\documentclass{iopjournal}
\usepackage{ragged2e}
\usepackage{titlesec}
\usepackage{hyperref}
\usepackage{soul}
\usepackage{xcolor}
\renewcommand{\hl}[1]{\textcolor{red}{#1}}
 \usepackage{comment}
\usepackage{amsmath,bm}
  \usepackage{caption}
    \usepackage{subcaption}
\justifying
\begin{document}
\articletype{Topical Review} 
\title{Exchange-Correlation Functionals in 2D Materials: Applications, Challenges, and Limitations}
\author{Ahsan Javed$^{1, 2}$\orcid{0009-0007-3657-4566}, Mahvish Shaheen$^1$, Muhammad Shahbaz$^3$\orcid{0000-0002-6611-2641}, M. Sufyan Ramzan$^4$\orcid{0000-0002-5017-8718}, Rafi Ullah$^5$\orcid{0000-0002-8468-4678}, Wei Jiang$^{1,*}$\orcid{0000-0001-6167-1156}}

\affil{$^1$School of Physics, Beijing Institute of Technology, Beijing, China}

\affil{$^2$Department of Physics, Lahore University of Management Sciences (LUMS), Lahore, Pakistan}

\affil{$^3$Department of Physics, University of the Punjab, Lahore, Pakistan}

\affil{$^4$Institut für Festkörpertheorie und Optik, Friedrich-Schiller-Universität Jena, Germany}

\affil{$^5$Department of Physics, Hiram College, 
Hiram OH 44234, USA.}

\affil{$^*$Author to whom any correspondence should be addressed.}

\email{wjiang@bit.edu.cn}
\\
\begin{abstract}
\begin{justify}
The rapid development of two-dimensional (2D) materials has reshaped modern nanoscience, offering properties that differ fundamentally from their bulk counterparts. As experimental discovery accelerates, the need for reliable computational techniques has become increasingly important. Within the framework of density functional theory, this review explores the critical role of exchange-correlation functionals in predicting key material properties such as structural, optoelectronic, magnetic, and thermal. We examine the challenges posed by quantum confinement, anisotropic screening, and van der Waals interactions, which conventional functionals often fail to describe. Advanced approaches, including meta-GGA, hybrid functionals, and many-body perturbation theory (e.g., GW and Bethe-Salpeter equation), are assessed for their improved accuracy in capturing electronic structure and excitonic effects. We further discuss the non-universality of functionals across different 2D material families and the emerging role of machine learning to enhance computational efficiency. Finally, the review outlines current limitations and emerging strategies, providing a roadmap for advancing exchange-correlation functionals and beyond, to enable the practical design and application of 2D materials.
\end{justify}
\end{abstract}
\section{Introduction}
The groundbreaking discovery of graphene by K. Novoselov and A. Geim in 2004 marked a pivotal moment in the exploration of two-dimensional (2D) materials \cite{singh1998physics, graphene_2007}. Graphene consists of a single layer of carbon atoms arranged in a hexagonal pattern and, despite being just one atom thick, exhibits exceptional mechanical strength \cite{ares2022recent}. Its band structure has a zero-band gap with charge carriers having zero effective mass, resulting in exceptionally high electron mobility \cite{graphene_2008, graphene_bg_engineering}. However, the absence of a bandgap limits its potential in semiconductor applications, leading to the search for other 2D materials with desirable electronic properties \cite{Ramzan2024, khan2025altermagnetism, maroof2025first}. 

\begin{figure}[h!]
   \centering
    \includegraphics[scale=0.25]{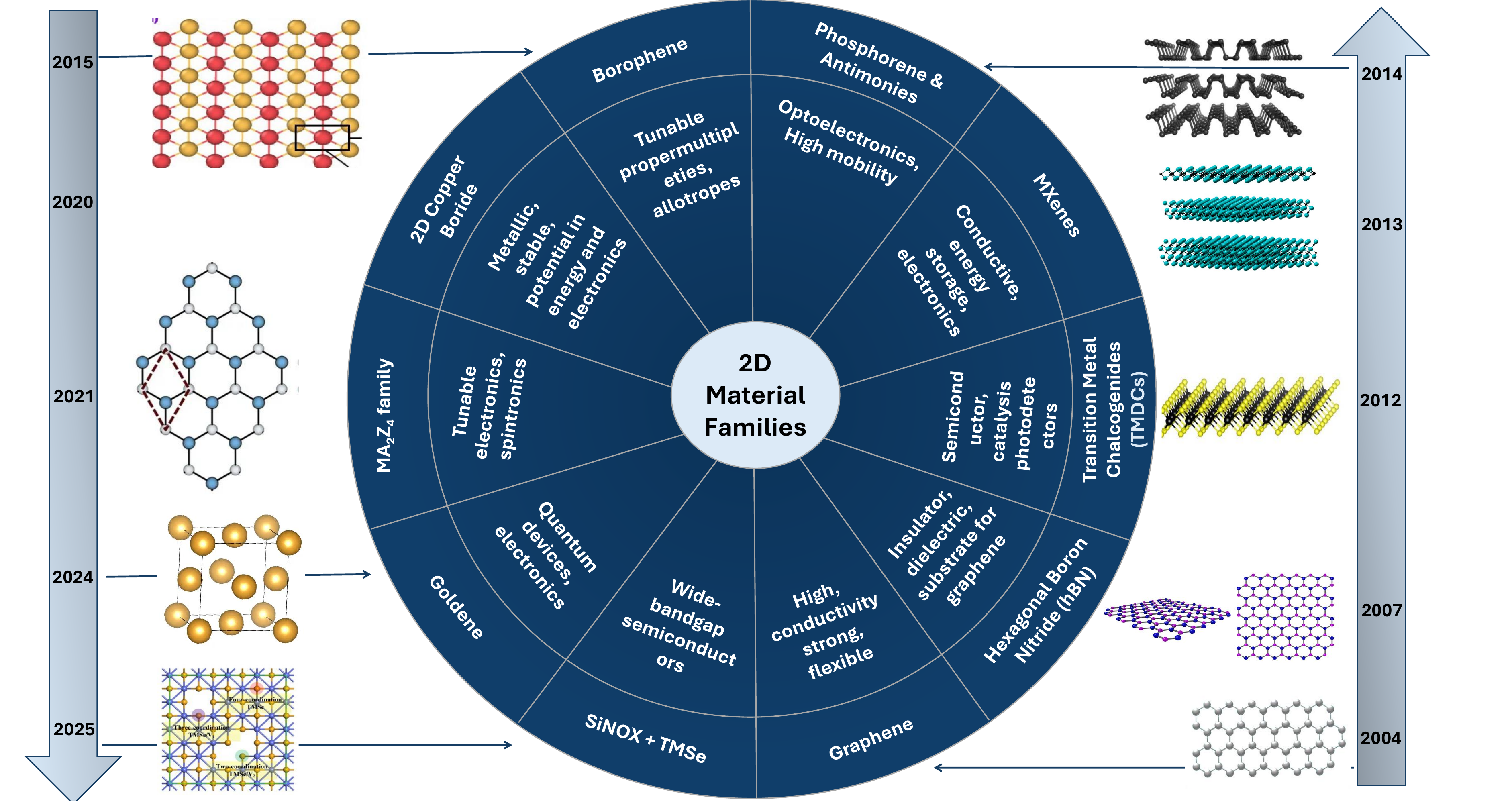}
    \caption{Families of 2D materials are represented by the outer layer, which shows the material family name, while the inner layer highlights their practical applications across various fields.}
    \label{fig:family}
\end{figure}

\noindent In the past decade, researchers have discovered a diverse family of 2D materials, including chalcogenides, oxides, hBN, and MXenes \cite{stehle2015synthesis, ali2021helicity, ramzan2019electronic, wozniak2023electronic}. They exhibit a wide range of electronic properties, ranging from insulating (e.g., hBN) \cite{sajid2018defect} to metallic (e.g., NbSe$_2$) \cite{lco_NbSe2_2020} and semiconducting (e.g., SiNOH) \cite{zhang2025theoretical}, with band gaps as large as 6 $eV$ \cite{bg_range_2020}. Transition metal dichalcogenides (TMDCs) show layer-dependent bandgaps i.e., transitioning from indirect gaps in bulk to direct gaps in monolayers. This makes them promising candidates for applications in photodetectors \cite{tmdcsphotodetectors, mustafa2023first, ahsan2026advances} and light-emitting devices \cite{wang2024integration, ismael2024two, sohail2024recent}. MXenes are well-suited for energy storage \cite{ti3_mXene, mustafa2023investigation, ghazal2026advanced}, while phosphorene exhibits remarkable hole mobility \cite{liu2014phosphorene}. The MnBi$_2$Te$_4$ monolayer, known for its intrinsic magnetic topological insulator properties, offers significant potential for quantum computing \cite{li2019intrinsic, vyazovskaya2025intrinsic}. Metallic 2D materials, such as NbSe$_2$, are being investigated for their potential in superconductivity \cite{saito2016highly, qiu2021recent}. Antiferromagnetic materials such as Cr$_2$O$_3$ and FePS$_3$ are being explored for their potential in low-power spintronic devices \cite{hashmi2020ising, wei2020emerging}, while 2D metal-organic frameworks (MOFs) are being investigated for applications in gas storage, sensing, and catalysis \cite{dhakshinamoorthy20192d, yuan2022metal, khan2023current} as well as exotic quantum states \cite{Wang2013prediction,Wang2013quantum,Jiang2019alieb,Jiang2020topological,Jiang2021exotic}. This rapidly expanding library of 2D materials has opened new avenues for electronic, optoelectronic, spintronic, and energy applications \cite{2D_npj_overview_2020, reviewTMDCS_2015, mXene_2024review, abdullah2026gaag, abdullah2026inag}, as summarized in Fig.~\ref{fig:family}. 
\noindent The reduced dimensionality of 2D materials leads to extraordinary properties due to quantum confinement effects but also presents significant theoretical challenges. Enhanced electron correlations, weak interlayer forces and strong spin-orbit effects require specialized computational methods to accurately predict material behavior. Standard computational frameworks often fail to capture these effects, resulting in discrepancies between theoretical predictions and experimental measurements \cite{ratcliff2017challenges, 2D_npj_overview_2020}. \\
\noindent Density functional theory (DFT) has emerged as a leading computational method in providing insights into a material's structure and its associated properties. At the core of DFT is the exchange-correlation (XC) functional, which approximates the complex electron interactions and determines the total energy of the system. The choice of XC functional plays a pivotal role in the accuracy of material property predictions, and this landscape has evolved from local density approximations to generalized gradient, hybrid, meta-GGA, and most recently, machine-learned functionals, each addressing different aspects of electronic exchange and correlation \cite{perdew_gga_1998, dft2015chemist,snyder2012finding,schleder2019dft,bystrom2024training}. \\
\noindent In this review, we begin with a brief theoretical background on XC functionals, followed by an exploration of the computational challenges posed by 2D materials. We then delve into the structural, opto-electronic, thermal, and magnetic properties of 2D materials, highlighting how different XC functionals perform in predicting these important characteristics, and systematically identifying where current functionals excel and where they fall short. Finally, we examine the role of machine learning, which is advancing the next generation of computational techniques by enhancing both accuracy and efficiency across various aspects of material modeling. Our goal is to provide a clear picture of the current state and suggest promising directions for future research. \\ 
\noindent Our work distinguishes itself from prior review articles, which have typically adopted a broader perspective. For example, Carvalho et al. (2021) explored computational methods for predicting optoelectronic properties in 2D materials, emphasizing the challenges of reduced dimensionality, but did not address meta-GGA or hybrid functionals \cite{carvalho2021computational}. Similarly, Penev et al. (2021) provided an extensive overview of the theoretical prediction of 2D materials, covering general modeling methods, structures, properties, device functionalities, and synthesis routes \cite{penev2021theoretical}. Gupta et al. (2025) focused solely on 2D transition metal dichalcogenides \cite{gupta2025two}. While these reviews provide valuable insights into the diverse aspects of 2D materials, none have directly focused on critically evaluating and comparing the performance of different approaches in achieving chemical accuracy.
\section{\hl{Theoretical framework}}
\begin{figure}[h!]
    \centering
    \includegraphics[scale=0.65]{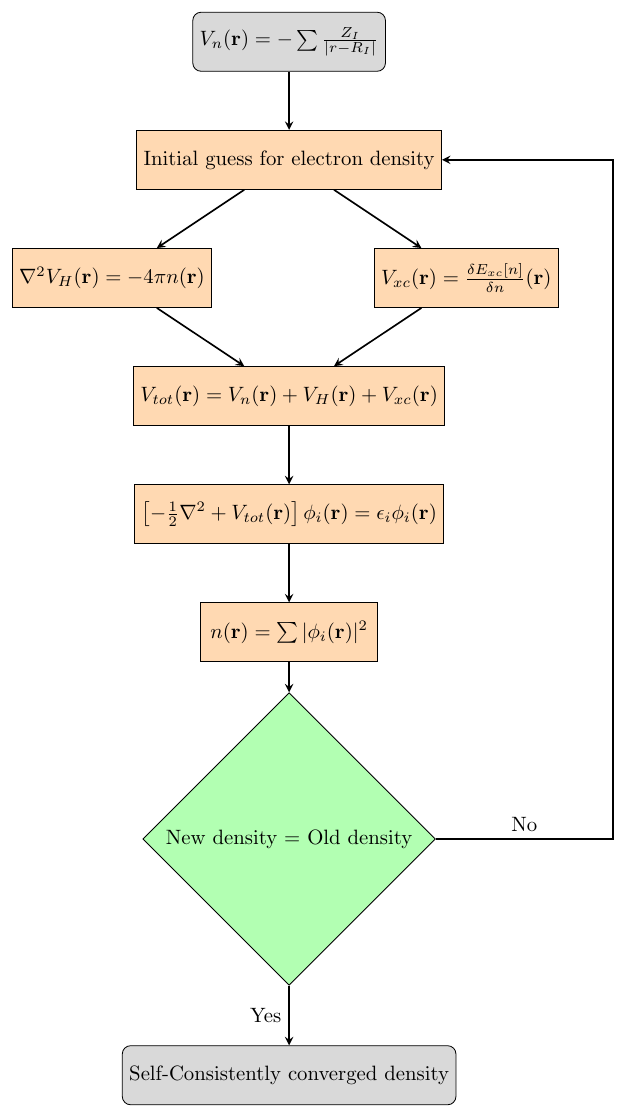}
    \caption{The schematic diagram of KS-DFT. Reprinted from \cite{javed2024band}, copyright (2024), with permission from Elsevier.}
      \label{fig:model}
\end{figure}

The behavior of materials at the quantum level is inherently complex, with interactions between electrons and nuclei forming a quantum many-body system. The total Hamiltonian that governs these interactions is:
\begin{equation}
 \hat{H} = - \sum_i \frac{\nabla_i^2}{2} - \sum_I \frac{\nabla_I^2}{2M_I} - \sum_{i,I} \frac{Z_I}{|\mathbf{r}_i-\mathbf{R}_I|}+\frac{1}{2}  \sum_{i \neq j}\frac{1}{|\mathbf{r}_i-\mathbf{r}_j|}+\frac{1}{2}\sum_{I \neq J} \frac{Z_I Z_J}{|\mathbf{R}_I-\mathbf{R}_J|},
\end{equation}
where, $\mathbf{r}_i$ is the position vector for the $i$th electron, and $\mathbf{R}_I$, $M_I$, and $Z_I$  are the position, mass and charge of the $I$th nucleus respectively. This Hamiltonian represents the total energy of a system composed of electrons and nuclei. The first term corresponds to the kinetic energy (K.E.) of the electrons, while the second term accounts for the K.E. of the nuclei. The third term denotes the Coulomb attraction between electrons and nuclei, reflecting the electrostatic forces that bind the system together. The fourth term describes the e–e Coulomb repulsion, which arises from the mutual repulsive interaction. The last term represents the nuclear–nuclear Coulomb repulsion, capturing the electrostatic repulsion between the positively charged nuclei. In principle, solving the Schrödinger equation ($\hat{H}\Psi = E_{tot}\Psi$) exactly would yield complete knowledge of the properties of a material. One could simplify the problem using the fact that electrons move significantly faster than nuclei and could adjust their distribution for any given nuclear configuration. Thus, we could reduce the problem to the motion of electrons in the nuclear potential of clamped nuclei. This is known as the Born-Oppenheimer approximation. However, the exponential complexity of electron–electron interactions makes such calculations infeasible for systems larger than just a few particles~\cite{giustino2014materials}. \\ 
\noindent DFT overcomes this challenge by reformulating the problem in terms of electron density $\rho(\mathbf{r})$ rather than the many-body wavefunction $\Psi$, drastically reducing computational complexity. The Hohenberg-Kohn (HK) theorems \cite{hohenberg1964inhomogeneous} provide the foundation for DFT and state the following:
\begin{itemize}
    \item The ground-state electron density uniquely determines all properties of the system,
    \item There exists a universal energy functional $E[\rho]$ that is minimized by the true ground-state density.
\end{itemize}
While the HK theorems guarantee the existence of a universal energy functional  $E[\rho]$ and provide a variational principle, its exact form remains unknown. Consequently, the central challenge of DFT is not whether to minimize a functional, but which approximate functional to use. The Kohn-Sham (KS) approach \cite{kohn1965self} addresses this by mapping the real interacting system onto an equivalent system of non-interacting electrons that reproduces the true electron density. This mapping leads to a set of effective one-electron equations whose accuracy depends critically on the exchange–correlation functional. The electron density in the KS system is expressed as:
\begin{equation}
\rho(\mathbf{r}) = \sum_{i=1}^{occ} |\psi_i(\mathbf{r})|^2
\end{equation}
where $\psi_i$ are the KS orbitals that satisfy the one-electron equations:

\begin{equation}
\left[-\frac{1}{2}\nabla^2 + V_{\rm eff}(\mathbf{r})\right]\psi_i(\mathbf{r}) = \epsilon_i \psi_i(\mathbf{r})
\end{equation}
The effective potential $V_{\rm eff}$ connects back to HK through:

\begin{equation}
V_{\rm eff}(\mathbf{r}) = V_{\rm ext}(\mathbf{r}) + \int \frac{\rho(\mathbf{r}')}{|\mathbf{r}-\mathbf{r}'|}d\mathbf{r}' + \frac{\delta E_{\rm xc}[\rho]}{\delta\rho(\mathbf{r})}
\end{equation}
Here, $V_{\rm eff}(\mathbf{r})$ represents the effective potential acting on an electron within the Kohn–Sham framework. $V_{\rm ext}(\mathbf{r})$ denotes the external potential, which normally originates from the electrostatic attraction between the electrons and the nuclei. The second term corresponds to the Hartree potential, describing the classical Coulomb repulsion arising from the electron density distribution. The final term represents the exchange–correlation potential, $V_{\rm xc} [\rho]=\frac{\delta E_{\rm xc}[\rho]}{\delta\rho(\mathbf{r})} $, where $E_{\rm xc} [\rho]$ is the exchange-correlation energy. It incorporates the many-body quantum mechanical effects of electrons, i.e., exchange and correlation, that go beyond classical electrostatic interactions. This mapping elegantly transforms the many-body problem into a computationally manageable framework~\cite{kohn1999nobel} as illustrated in Fig.~\ref{fig:model}. However, the accuracy of the KS equations depends critically on the exchange-correlation potential $V_{\rm xc} [\rho]$ despite being the smallest part of $V_{\rm eff}(\mathbf{r})$. Since the exact form of the XC functional is unknown, various approximation levels have been developed, each balancing accuracy and computational cost, forming the basis of Jacob’s ladder of DFT (Fig. \ref{fig:jacob}).
\subsection{Exchange-Correlation functionals}

\noindent Local density approximation (LDA) is one of the simplest and most commonly used methods in DFT. It assumes that the exchange–correlation energy of an electron system at each point in space can be approximated by that of a uniform electron gas having the same local density (see Eq. 5). This approach provides good accuracy for systems that are nearly homogeneous, but its performance deteriorates in regions where the electron density varies strongly~\cite{perdew1992accurate_lda}. Generalized gradient approximation (GGA) improves accuracy by including the gradient of the electron density to approximate the exchange–correlation energy (Eq. 6), but still struggles with complex electron correlations \cite{perdew_gga_1998}. \hl{Local and semi-local approximations (LDA/GGA) suffer from three key shortcomings: a short-range exchange-correlation hole, self-interaction error, and a lack of derivative discontinuity with respect to the number of particles. The first flaw prevents the description of long-range electron correlations (van der Waals interactions)}~\cite{shahbaz2018semilocal}. \hl{The latter two are the fundamental reasons for the notorious underestimation of band gaps in semiconductors and insulators. In 2D materials, these limitations lead to underestimated band gaps, inaccurate lattice constants, and poor treatment of anisotropic electronic environments, driving the need for more advanced functionals} \cite{javed2025structural}. \\ 

\noindent To address these issues, meta-GGA functionals were developed, which include not only the density and its gradient but also the kinetic energy density $\tau = \sum_i^{occ} \frac{|\bm\nabla \psi_i|^2}{2}$ (Eq. 7), providing a more accurate description of the XC functional. By incorporating a more non-local description, these advanced functionals provide a longer-ranged exchange-correlation hole, significantly mitigate self-interaction error, and can even replicate the crucial derivative discontinuity. \\
\noindent The exchange correlation energy in LDA, GGA, and meta-GGA approximations could be expressed as follows:
 \begin{align}
    E_{\rm xc}^{\rm LDA}[\rho] &= \int \rho(\mathbf{r}) \, \epsilon_{\rm xc}^{\rm LDA}\big(\rho(\mathbf{r})\big)  \, d\mathbf{r}, \\
    E_{\rm xc}^{\rm GGA}[\rho] &= \int \rho(\mathbf{r}) \, \epsilon_{\rm xc}^{\rm GGA}\big(\rho(\mathbf{r}), \nabla \rho(\mathbf{r})\big)  \, d\mathbf{r},
\end{align}
\begin{align}
E_{\rm xc}^{\rm meta\text{-}GGA}[\rho] &= \int \rho(\mathbf{r}) \, \epsilon_{\rm xc}^{\rm meta\text{-}GGA}\big(\rho(\mathbf{r}), \nabla \rho(\mathbf{r}), \tau(\mathbf{r})\big)  \, d\mathbf{r},
\end{align}
where $\rho(\mathbf{r})$ is the electron density, $\nabla \rho(\mathbf{r})$ its gradient and $\tau(\mathbf{r})$ is the kinetic energy density. $\epsilon_{\rm xc}^{\rm X}$ (X=LDA.GGA,meta-GGA) is the XC energy per particle in each of the approaches. \\

\noindent \hl{DFT suffers from several well-known limitations. Among these are the self-interaction error and the absence of the derivative discontinuity in the XC energy, which can lead to un-physical charge delocalization} \cite{perdew1983physical, cohen2008insights}. \hl{These deficiencies manifest as the underestimation of band gaps and, consequently, errors in the prediction of optical properties} \cite{martin2020electronic}. \hl{For 2D materials, the self-interaction error is particularly pronounced due to reduced dielectric screening, which enhances electron-electron interactions} \cite{mueller2018exciton}. \hl{Another major limitation arises from the short-range nature of the XC hole in standard DFT functionals, which fails to capture long-range electron correlations such as van der Waals (vdW) interactions} \cite{shahbaz2018semilocal}. \hl{These interactions are especially significant in highly polarizable 2D materials, such as graphene, and in layered systems including vdW heterostructures, where interlayer binding is dominated by dispersion forces} \cite{geim2013van, oh2024review}. \hl{Therefore, incorporating corrections for missing vdW interactions is essential for a reliable description of such 2D systems.} 

\noindent \hl{Typical LDA and GGA functionals suffer from all of the problems described above. Higher‑level approximations are designed to overcome these limitations. Some meta‑GGA functionals, such as TASK} \cite{aschebrock2019ultranonlocality}, \hl{have been constructed with enhanced nonlocality to yield improved band gaps, and subsequent modifications like mTASK further refine this performance for low‑dimensional materials} \cite{neupane2021opening}. \hl{More recently, the Lebeda–Aschebrock–Kümmel (\textsc{LAK}) meta‑GGA has been proposed to balance accurate band gaps and energetics within a semilocal framework} \cite{lebeda2025meta}. \hl{Hybrid functionals like HSE06 similarly improve band gap predictions by incorporating exact exchange.} 

\noindent \hl{The shortcomings related to long‑range correlations can be addressed through nonlocal correlation functionals, such as vdW‑DF2} \cite{Lee:10}, \hl{vdW‑DF3} \cite{chakraborty2020next}, \hl{VV10} \cite{VV:10}, \hl{and rVV10} \cite{rvv10}, \hl{or via empirical, semi‑empirical dispersion corrections, including D3} \cite{grimme2010consistent}, \hl{D4} \cite{caldeweyher2019extension}, \hl{and many‑body dispersion methods} \cite{tkatchenko2012accurate}. \hl{These approaches are crucial for accurately capturing interlayer binding and other dispersion‑dominated properties in 2D materials} \cite{pushkarev2023nature, tang2022high} \\
\noindent The SCAN (Strongly Constrained and Appropriately Normed) functional is a prominent example of a meta-GGA functional, as it obeys the seventeen exact physical constraints and has proven effective for a wide range of materials. Its key strength lies in a significant reduction of self-interaction error and accurate inclusion of non-local interactions and strong electron correlations, critical for understanding low-dimensional systems~\cite{sun2015scan, perdew2025scan}. SCAN accounts for short-range vdW interactions, while long-range vdW effects may be incorporated via non-local correlation functionals. Several of these methods have been evaluated in Ref.~\cite{shahbaz2019evaluation}. The role of nonlocal correlations (vdW interactions) in structural, dynamic, electronic, and superconducting properties of materials has been recently studied in Refs.~\cite{bakar2023effects, Azam:24, Rehman_2024, Bakar:PdCPtC25}. Typically, rVV10 functional (revised Vydrov–Van Voorhis nonlocal functional)~\cite{rvv10} is used for this purpose due to its consistent accuracy, especially in the long-range region. The inclusion of vdW interactions enhances the accuracy in structural and optoelectronic properties~\cite{rvv10_peng2015scan+}. The rVV10 functional involves the modification of the original VV10 functional to enable efficient integration for nonlocal correlation in the plane wave frame-work in addition to refinement of an empirical parameter. The SCAN+rVV10 XC energy could be written as
\begin{equation}
    E_{\text{xc}}^{\text{SCAN+rVV10}}[\rho] = E_{\text{xc}}^{\text{SCAN}}[\rho] + E_{\text{vdW}}^{\text{rVV10}}[\rho].
\end{equation} 
\noindent \hl{Despite these advances, semi-local and meta-GGA functionals often underestimate electronic band gaps mainly because they treat exchange interactions only approximately. Conversely, the Hartree Fock (HF) method tends to overestimate band gaps because it neglects correlation entirely}~\cite{franchini2015electronic}. \hl{The adiabatic connection formula suggests that the exchange-correlation energy should include HF exchange partially due to the coupling-strength integration}~\cite{Becke_Hybrid:93,Perdew_Hybrid:96}. \hl{The hybrid functionals were developed, mixing a fraction of exact HF exchange with DFT exchange-correlation, leading to an increase in the band gaps. This approach improves accuracy but comes with a higher computational cost, which can be restrictive for large-scale simulations.}
\begin{equation}
E_{\rm xc}^{\rm hybrid} = \alpha E_{\rm x}^{\rm HF} + (1-\alpha) E_{\rm x}^{\rm DFT} + E_{\rm c}^{\rm DFT},
\end{equation}
where $E_{\rm x}^{\rm HF}$ is the exact HF exchange, $E_{\rm x}^{\rm DFT}$ and  $E_{\rm c}^{\rm DFT}$ are the DFT exchange and correlation terms, and $\alpha$ is the mixing parameter.  
HSE06 is a widely adopted screened hybrid functional. By refining the treatment of exchange interactions it provides a good compromise between accuracy and computational efficiency, making it a standard choice for calculating electronic structures and band gaps in both bulk and low-dimensional materials \cite{hybrid_2011}.
    \begin{equation}
        E^{\rm HSE06}_{\rm xc} = \frac{1}{4} E^{\rm HF,SR}_{\rm x} + \frac{3}{4} E^{\rm PBE,SR}_{\rm x} + E^{\rm PBE,LR}_{\rm x} + E^{\rm PBE}_{\rm c}
    \end{equation}
\subsection{GW+BSE framework}
The GW approximation, a method for calculating electronic properties, goes beyond the mean-field, independent-particle framework of DFT \cite{shishkin2007accurate, hybertsen1986electron}. \hl{By accounting for many-body electron-electron interactions, GW provides a more accurate representation of these interactions, leading to improved predictions of properties such as band gaps and excitation energies for low-dimensional materials} \cite{PhysRevB.74.035101, PhysRevB.75.235102}. Here $G$ stands for the Green's function, which represents the propagation of an electron within a material and is given as,

\begin{equation}
    G(k, \omega) = [\omega - H(k) - \Sigma(k, \omega)]^{-1}
\end{equation}
$G(k, \omega)$ is the Green's function at wave vector $k$ and energy $\omega$, $H(k)$ is the one-electron Hamiltonian which contains the kinetic energy and effective potential due to electron-electron interactions. $\Sigma(k, \omega)$ is the self-energy, which represents the many-body effects, given by Dyson equation, 
\begin{equation}
    \Sigma(k, \omega) = iG(k, \omega)^{-1} - [iG^0(k, \omega)^{-1} - \Sigma_{xc}(k, \omega)]^{-1}
\end{equation}
where $G^0(k, \omega)$ is the non-interacting Green's function, often calculated from the one-electron Hamiltonian, $\Sigma_{xc}(k, \omega)$ represents the exchange-correlation self-energy, accounting for effects beyond the Hartree term in the electron-electron interaction. Coulomb interaction ($W$) is given as,
\begin{equation}
    W(q, \omega) = v(q) + v(q) \cdot \epsilon(q, \omega) \cdot W(q, \omega)
\end{equation}
where $v(q)$ is the bare Coulomb interaction, typically given by the $\frac{1}{r}$ form of the Coulomb potential, $\epsilon(q, \omega)$ is the dielectric function, describing the material's response to an external electric field \cite{gw_2019}. The GW method involves self-consistently solving these equations for $G$, $W$, and $\Sigma$ to obtain an improved description of the electronic structure and properties of a material, particularly in terms of electronic excitations, band gaps, and optical properties. The self-consistency arises because G and $\Sigma$ depend on each other, and the equations are iteratively solved until convergence is achieved. \\
\noindent Bethe-Salpeter equation (BSE) is used to incorporate excitonic effects \cite{PhysRevLett.80.4510, PhysRevLett.81.2312} and solved following the GW calculation in order to include electron-hole interactions. The BSE is typically expressed as,
\begin{equation*}
\resizebox{\textwidth}{!}{
($E_{cv}^{BSE} - E_g)A_{cv}({k}, {k'}) = \sum_{c'v'}\int d^3r d^3r' \Psi_{c'v'}^{cv}({r}, {r'}) \times [W({r}, {r'}) - \sum_{c''v''} V({r}, {r''}) \chi_{c''v''}^{cv}({r''}, {r'})]A_{c'v'}({r'}, {r})$}
\end{equation*}
\begin{itemize}
    \item $E_{cv}^{BSE}$ represents the exciton binding energy, which quantifies the energy required to create an electron-hole pair in an excited state compared to the ground state of the material,
    item $E_g$ is the fundamental band gap,
    \item $A_{cv}$ describes the amplitude of an electron-hole pair,
    \item Spatial distribution of the electron and hole is given by excitonic wave-function $\Psi_{c'v'}^{cv}({r}, {r'})$.
    \item $W$ and $V$ represents the direct Coulomb interaction and screened Coulomb interaction respectively between the electron and hole,
    \item $\chi_{c''v''}^{cv}({r''}, {r'})$ represents the dielectric susceptibility, describing how the material responds to the presence of an electron-hole pair \cite{gw_bse}.
\end{itemize}
BSE is solved to determine exciton binding energy and excitonic wave functions, is essential for accurately describing the optical properties. This approach allows for detailed predictions of excited states and optical spectra by providing insights into the interactions between electrons and holes. Such interactions lead to phenomena such as exciton absorption and emission, thereby enhancing our understanding of materials under optical excitation \cite{PhysRevResearch.2.032019, PhysRevB.92.045209}. \hl{Recent GW-BSE calculations on quantum-confined chiral halide perovskites further demonstrate that excitonic transitions can govern circular dichroism and chiroptical responses in low-dimensional materials}~\cite{shaheen2026circular,li2024large_cd_perovskite}.
\subsection{Jacob's ladder}
\begin{figure}[h!]
    \centering
    \includegraphics[scale=0.42]{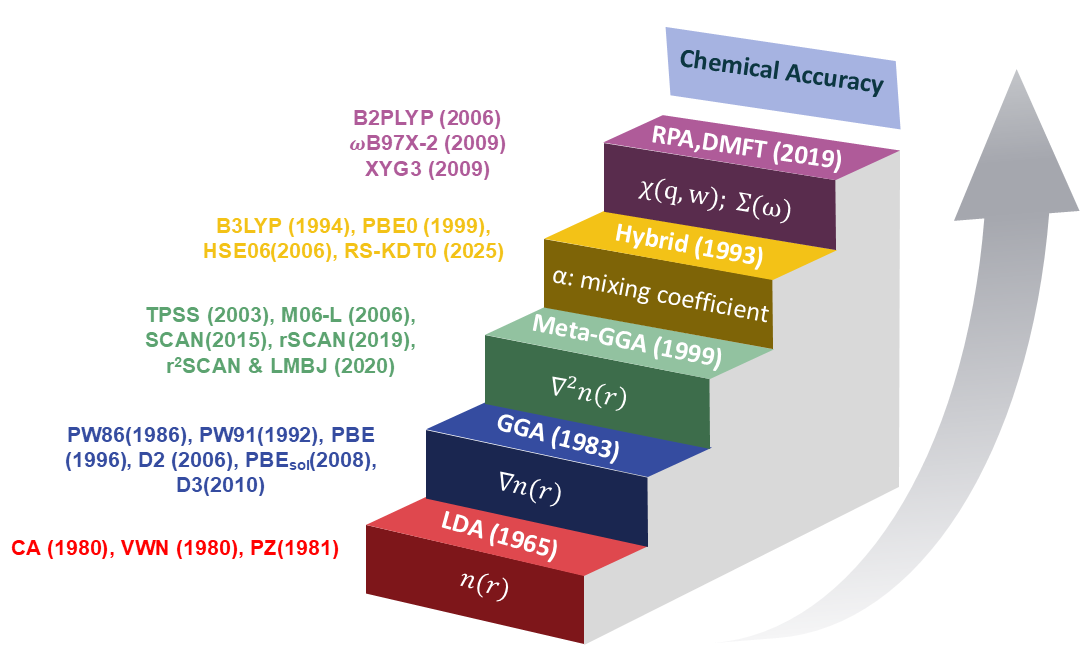}
    \caption{Jacob’s Ladder in DFT: A hierarchy of XC approximations where each higher rung offers improved accuracy. On the left, the famous functionals along with their years of publication are listed, showcasing the evolution of XC functionals from LDA to hybrid and double-hybrid methods.}
    \label{fig:jacob}
\end{figure}

Jacob’s ladder provides a conceptual framework for organizing XC functionals based on the level of physical information included and their corresponding accuracy as shown in Fig. \ref{fig:jacob}. Each rung of the ladder represents a step toward ``chemical accuracy'', starting from simple local density approximations at the bottom and progressing toward more sophisticated hybrid and many-body methods at the top. The five rungs consist of LDA, GGA, meta-GGA, hybrid functionals mixing exact and approximate exchange, and fully nonlocal or orbital-dependent methods such as Random phase approximation (RPA), Dynamical mean-field theory (DMFT) or GW respectively. Higher rungs typically give qualitatively more accurate trends, but at a higher computational cost \cite{car2016fixing}.
\newpage
\noindent \hl{For 2D materials, the Jacob’s ladder hierarchy becomes particularly critical. The reduced dimensionality often leads to weaker dielectric screening and stronger quantum confinement, which amplify exchange-correlation effects and render lower-rung functionals (e.g., LDA and GGA) often inadequate for predicting key properties such as band gaps, exciton binding energies, and interlayer interactions. Climbing to higher-rung functionals, such as Hybrid functionals and meta-GGA, offer improved descriptions of these effects and partially account for van der Waals interactions. Nevertheless, accurate treatment of excitonic phenomena generally requires many-body approaches like GW+BSE. Given the substantial computational cost of these methods, their application in large-scale and high-throughput studies of 2D materials remains challenging. Thus, balancing accuracy and efficiency remains a central issue, particularly for layered materials under strain, external fields, or different stacking configurations} \cite{ghaebi2025tunable, ramzan2025strain}.
\section{Computational challenges in modeling 2D materials}
DFT has greatly advanced our understanding of electronic structures, especially for bulk materials. But for 2D materials, DFT shows limitations due to their unique properties. The reduced dimensionality, strong many-body interactions, and excitonic effects lead to behaviors that differ greatly from bulk systems. To model these systems accurately, improved functionals and advanced methods beyond standard DFT are necessary
\subsection{Quantum confinement and Anisotropic Dielectric screening}
One of the key challenges in 2D materials arises from quantum confinement, where electrons are restricted to a single plane, leading to enhanced electron–electron interactions and stronger correlation effects compared to bulk systems. Standard XC functionals often fail to accurately capture this enhancement \cite{li2024charge, cipriano2020quantum}. In addition, the dielectric environment in 2D materials is highly anisotropic: while electron density is confined in-plane, the out-of-plane direction interfaces with vacuum, resulting in direction-dependent screening. However, traditional semilocal functionals assume isotropic screening, which is an oversimplification for such systems and leads to underestimated band gaps and related inaccuracies in electronic properties \cite{lorke2025anisotropic, rudenko2024anisotropic, latini2015excitons}. \\
\noindent \hl{For example, the indirect band gap of bulk WS$_2$ ($\sim$1.3 eV) transitions to a direct band gap of $\sim$2.0 $eV$ in the monolayer} \cite{NGUYEN2025181602}. \hl{While LDA predicts 1.67 $eV$ and PBE 1.58 $eV$, the HSE06 functional yields $\sim$2.1 $eV$, whereas GW calculations provide quasiparticle (QP) band gaps of $\sim$2.5 eV} \cite{javed2025structural}. \hl{A broader comparison is provided in Table 3.} \\
\noindent \hl{Exciton binding energies (EBE) in 2D materials range from 0.3-2.0 $eV$ across TMDCs and other 2D materials} \cite{mueller2018exciton, javed2025investigation}, \hl{far exceeding those in bulk counterparts. This enhancement arises from reduced dimensionality and strong electron-hole correlation. However, local and semi-local XC functionals fail to capture the crucial nonlocal electron–hole interactions. Consequently, many body methods, such as the GW method combined with BSE or quantum Monte Carlo approaches, are required} \hl{for accurate excitonic spectra} \cite{Ashwin_2012}. \hl{These methods, while essential for precision, impose significant computational costs, limiting the scale and complexity of feasible systems} \cite{gw_bse, javed2025machine}.
\subsection{Van der Waals interactions}
Another major challenge in 2D materials is the accurate description of vdW interactions, which govern interlayer binding in these systems. Although significantly weaker than covalent bonds, vdW forces are critical for determining structural stability, stacking order, and interlayer spacing. Semilocal XC functionals typically underestimate these interactions, leading to inaccurate predictions of stacking energetics and equilibrium geometries in heterostructures. To address this, non-local correlation functionals, such as rVV10 and vdW-DF, explicitly incorporate long-range dispersion effects \cite{rvv10_peng2015scan+}. Despite these improvements, resolving small energy differences between competing stacking configurations remains challenging, particularly in heterostructures, where slight variations in interlayer coupling can significantly impact electronic properties \cite{stohr2019theory, momeni2020multiscale}. \hl{High-throughput vdW-DF2 calculations across over 230 layered 2D materials reveal that more than 80$\%$ of systems have interlayer binding energies between -10 $meV$\AA$^{-2}$ and -20 $meV$ \AA$^{-2}$, with equilibrium separations around 3 - 4 \AA. In the case of MoS$_2$, vdW-DF2 predicts an interlayer spacing of $\sim$3.8 \AA\ with a binding energy of -14.68 $meV$ \AA$^{-2}$, while optB88-vdW tends to overestimate the binding energies} \cite{tang2022high}.
\subsection{Spin-orbit coupling (SOC) and Excitonic effects}
Beyond quantum confinement and dispersion forces, certain classes of 2D materials (particularly TMDCs), exhibit strong SOC due to the presence of heavy transition metals like W and Mo, which enhances relativistic interactions. This leads to valence band splitting (VBS) and poses significant challenges for standard XC approximations \cite{valley}. \hl{For example, the band gaps of WSe$_2$ and WS$_2$ monolayers are 1.55 $eV$ and 1.82 $eV$, respectively, when SOC is not included. However, incorporating SOC slightly reduces the band gap to 1.22 $eV$ and 1.58 $eV$ (Table 3), demonstrating the impact of SOC on the electronic structure. This reduction in the band gap arises from spin splitting of the valence band, which modifies the overall band structure. The extent of VBS in these materials varies with the level of approximation used. For instance, the experimental VBS for the WSe$_2$ monolayer is approximately 0.52 eV} \cite{SOC_wse2}, \hl{while PBE predicts 0.43 $eV$, HSE06 estimates 0.58 $eV$, and G$_0$W$_0$ provides 0.54 eV} \cite{javed2025investigation}. \hl{The experimental VBS is in closest agreement with the G$_0$W$_0$ calculations, which is consistent for other 2D materials as well} \cite{javed2025investigation, SOC_ws2_1, SOC_ws2_2}. \\
\subsection{Non-universality of XC functional performance}
A further complication arises from the non-universality of XC functional performance across different 2D material families. \hl{For instance, PBE and LDA functionals tend to underperform for TMDCs compared to materials like graphene or phosphorene} \cite{patra2021efficient, radial_mae}. \hl{HSE06, while accurate for some systems, artificially opens a small gap in gapless 2D materials like graphene and silicene (Figure 10). The accuracy of a functional varies depending on the material property being predicted, i.e., PBE, for instance, correctly validates Raman peaks but performs poorly for band gap crossover studies in monolayer MoS$_2$} \cite{ali2024room}. \hl{Ensuring consistent accuracy across different material types and properties remains a central challenge in the field.}
\section{Evaluating the performance of XC functionals in 2D materials}
\subsection{Structural properties}
The lattice constant represents the physical dimension of the unit cell in a crystal structure. \hl{In conventional DFT calculations, lattice constant optimization for 2D materials often deviates from experimental values due to quantum confinement and reduced dimensionality. This discrepancy can introduce errors in structure-dependent properties. For example, the experimental lattice constant of the WSe$_2$ monolayer is 3.28\AA, while the PBE-optimized value in the Computational 2D Materials Database (C2DB) is 3.32\AA, with all other properties in the database based on this result.} Given that the C2DB is widely used for machine learning models in material informatics \cite{c2db_2015}, the accuracy of lattice constants in such databases is critical for high-throughput screening. \hl{Therefore, functionals that yield lattice constants closer to experimental values are essential for reliable and physically meaningful predictions in computational materials science.}
\subsubsection{\hl{Performance of semilocal functionals}}
\begin{figure}[h!]
    \centering
    \includegraphics[scale=0.8]{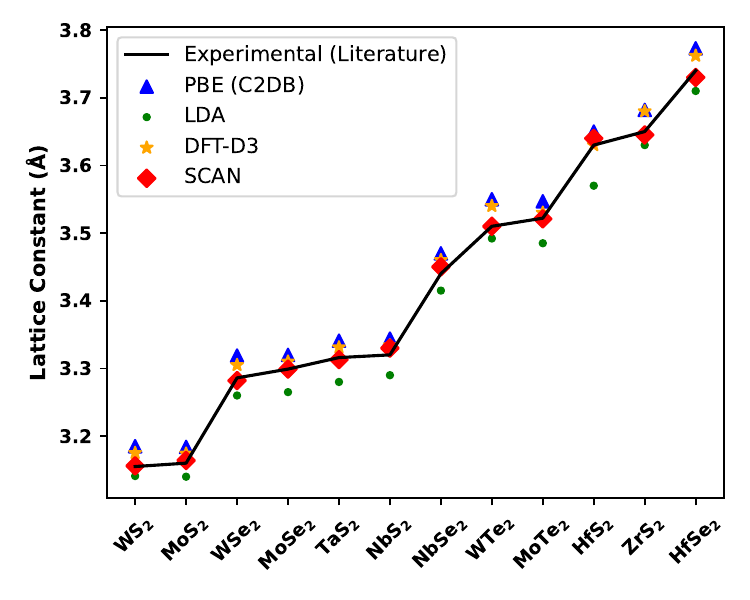}
    \caption{Comparison of optimized lattice constants for TMDC monolayers using different functionals, showing that the SCAN functional provides lattice constants closest to experimental values. Reprinted from \cite{javed2025structural}, copyright (2025), with permission from Elsevier.}
    \label{fig:lco}
\end{figure}

Early DFT studies using LDA and GGA optimized monolayer structures. Ding et al. calculated the structural properties of MX$_2$ monolayers (M=Mo, Nb, W, Ta; X=S, Se, Te) using LDA and PBE functionals \cite{ding2011first}, while Ataca et al. examined the stability of various monolayer compounds using LDA \cite{ataca2012stable}. These studies established a large library of 2D materials but revealed discrepancies in the predicted lattice parameters. They also observed differences in bond lengths: PBE tends to overestimate bond lengths and interlayer distances, while LDA underestimates them \cite{ding2011first, ataca2012stable}. Recent work by Yamusa et al., using GGA-PW91 and vdW+DF2 functionals for the MoS$_2$ monolayer, found that these functionals also overestimate the experimental lattice parameter \cite{yamusa2022elucidating}. \hl{This highlights the difficulty of accurately predicting lattice constants with semi-local functionals, primarily due to their inability to capture nonlocal exchange effects, which are significant in 2D materials. The discrepancies between predicted and experimental lattice constants for TMDC monolayers, shown in Fig.} \ref{fig:lco}, \hl{demonstrate the unreliability of semi-local functionals for lattice constants.} To predict structural properties accurately, dispersion corrections or higher-order XC functionals are essential, as discussed in the following section.
 \\
\subsubsection{\hl{Meta‑GGA functionals}}
\begin{figure}[h!]
\centering
    \includegraphics[scale=1.5]{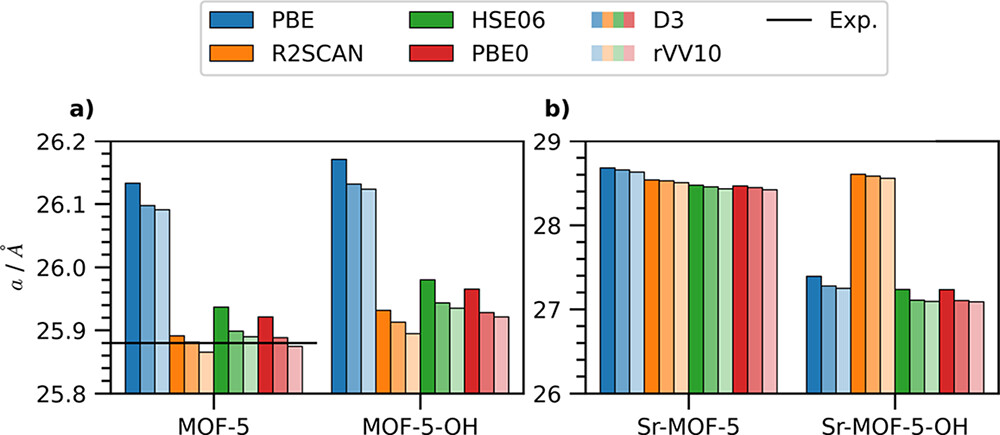}
    \caption{Lattice parameter $a$ of (a) conventional MOF-5 and (b) its Sr-substituted counterpart, featuring H-passivated (left) and hydroxyl-functionalized (right). The experimental reference for conventional MOF-5 \cite{li1999design} is marked by a horizontal bar. Reprinted from \cite{edzards2025benchmarking_MOF}. \href{https://creativecommons.org/licenses/by/4.0/}{\underline{CC BY 4.0}}.}
    \label{fig:MOF-5}
\end{figure}

\hl{These limitations in lattice constant prediction for 2D materials are being addressed by emerging meta-GGA functionals, which incorporate kinetic energy density along with electron density and its gradient, offering a more refined description of electron interactions.} SCAN, a meta-GGA functional, predicts lattice constants that closely match experimental values for a wide range of 2D materials \cite{javed2025structural, buda2017characterization}. This functional has shown particular effectiveness for TMDC monolayers (Fig. \ref{fig:lco}) and several other 2D systems (Table \ref{table:lco}). \hl{For example, PBE gives a lattice constant of 3.55 \AA\ for WTe$_2$, showing a significant deviation from the experimental value of 3.51 \AA, whereas SCAN predicts 3.51 \AA, matching the experiment.} However, many material families remain systematically untested, underscoring the need for further benchmarking.
\begin{table}[h!]
\setlength{\arrayrulewidth}{0.35mm}
\setlength{\tabcolsep}{8.5pt}
\renewcommand{\arraystretch}{1}
\begin{center}
\caption{\hl{Optimized lattice constants (\AA) for different 2D materials across different functionals, comparing predicted values with experimental data.}}
\label{table:lco}
\begin{tabular}{ |c|c|c|c|c|c| } 
\hline \hline
 & \textbf{LDA} & \textbf{PBE \cite{c2db_recent_2021}} & \textbf{DFT-D3} & \textbf{SCAN} & \textbf{\hl{Experiment}  } \\ \hline
Graphene & 2.445 & 2.470 & 2.510 & 2.461 & 2.460 \cite{lco_graphene} \\ \hline
hBN & 2.490 & 2.510 & 2.510 & 2.499 & 2.500 \cite{lco_hbn_2.5, lco_hbn_2.504} \\ \hline
WS$_2$ & 3.141 & 3.185 & 3.175 & 3.156 & 3.155 \cite{lc_ws2, lc_wse2}\\ \hline
MoS$_2$ & 3.140 & 3.184 & 3.174 & 3.164 & 3.160 \cite{lco_mos2_2016, lc_mose2_mos2_2019} \\ \hline
WSe$_2$ & 3.260 & 3.319 & 3.305 & 3.282 & 3.286 \cite{lc_wse2_2, lc_wse2} \\ \hline
MoSe$_2$ & 3.265 & 3.320 & 3.310 & 3.299 & 3.299 \cite{lc_mose2_mos2_2019} \\ \hline
TaS$_2$ & 3.280 & 3.341 & 3.331 & 3.313 & 3.316 \cite{lc_tas2} \\ \hline \
NbS$_2$ & 3.290 & 3.344 & 3.334 & 3.330 & 3.320 \cite{lco_NbS2_2012} \\ \hline
NbSe$_2$ & 3.415 & 3.470 & 3.460 & 3.450 & 3.440 \cite{lco_NbSe2_2020} \\ \hline
WTe$_2$ & 3.492 & 3.550 & 3.540 & 3.510 & 3.510 \cite{lco_WTe2_2016} \\ \hline
MoTe$_2$ & 3.485 & 3.547 & 3.530 & 3.521 & 3.522 \cite{lco_MoTe2_2023} \\ \hline
HfS$_2$ & 3.570 & 3.650 & 3.630 & 3.640 & 3.630 \cite{lco_HfS2_HfSe2} \\ \hline
ZrS$_2$ & 3.630 & 3.682 & 3.680 & 3.645 & 3.650 \cite{lc_ZrS2} \\ \hline
HfSe$_2$ & 3.710 & 3.773 & 3.762 & 3.730 & 3.740 \cite{lco_HfS2_HfSe2} \\
\hline
\end{tabular}
\end{center} 
\end{table} \\
\noindent In a recent study, Edzards et al. (2025) carried out a detailed benchmark of MOF‑5 and related structures using multiple XC functionals. Their results show that the meta-GGA functional r$^2$SCAN, combined with rVV10 dispersion corrections, provides the best balance between computational cost and accuracy \cite{edzards2025benchmarking_MOF}. The experimental lattice constant of MOF‑5 (a = 25.870 \AA) is reproduced almost exactly by r$^2$SCAN + rVV10, which gives 25.865 \AA, whereas PBE overestimates (26.133 \AA). This trend, illustrated in Fig. \ref{fig:MOF-5}, highlights the superiority of using a meta-GGA functional with explicit dispersion corrections for reliable structural predictions. Although MOF-5 is not a 2D material, its benchmarking demonstrates how meta-GGA functionals can reliably reproduce experimental lattice constants, making them strong candidates for further research.
\newpage
\subsubsection{Charge density waves}
Charge density waves (CDWs) are characterized by periodic modulations in the electron density, often accompanied by lattice distortions \cite{carpinelli1996direct}. These phenomena play a crucial role in superconductivity and phase transitions in metallic materials like monolayer TiSe$_2$, monolayer NbSe$_2$, and others. In 2D systems, the reduced dimensionality amplifies the importance of electron-phonon coupling, making accurate DFT modeling essential for predicting the stability and properties of CDWs \cite{xi2015strongly, chowdhury2022computational}. \\
\noindent Semilocal functionals are commonly used to model CDWs, but they often fail to capture the strong electron-electron correlations and electron-phonon coupling in low-dimensional systems. \hl{For example, Wines et al. (2025) showed that for monolayer 1T-VSe$_2$, PBE+U functionals with $U=1$ and 2 eV can reproduce the geometry reasonably well but fail to quantitatively capture the energy differences among magnetic and CDW states} \cite{cdw_scan_qm_vse2}. \\
\noindent SCAN has shown improved predictions for lattice constants (Section 4.1.2), which are crucial for accurately describing CDWs. \hl{For example, in a recent study by Li Yin et al., several meta-GGA functionals, including MVS, SCAN, r$^2$SCAN, TASK, and mTASK, were applied to monolayer TiSe$_2$. Their analysis of CDWs led to the recommendation of using SCAN and r$^2$SCAN for TMDC monolayers} \cite{cdw_scan_tise2}. \hl{These functionals help reduce self-interaction errors inherent in typical GGAs and provide a better description of correlation effects} \cite{cdw_scan_qm_vse2, cdw_scan_tise2}. r$^2$SCAN is a regularized version of SCAN designed to enhance numerical stability without compromising accuracy. r$^2$SCAN achieves this by relaxing some the exact constraints of SCAN, while maintaining results similar to those of SCAN \cite{furness2020accurate}. \\ 
\noindent In another study, DFT combined with Quantum Monte Carlo was utilized to benchmark several XC functionals for 1T-VSe$_2$, where SCAN and r$^2$SCAN provided \hl{accurate results for transition temperatures in both undistorted (228$K$) and CDW-distorted phase (68$K$)} \cite{cdw_scan_qm_vse2}. Wei-Chi Chiu et al. (2025) demonstrated that SCAN calculations correctly predict the nonmagnetic ground state of the CDW phase in NbSe$_2$ monolayer, consistent with previous research showing that CDW formation suppresses magnetic instability \cite{cdw_nbse2_strain}, further validating SCAN’s effectiveness in modeling CDWs in 2D materials.
\subsubsection{Ferro-electricity}
\noindent Ferroelectricity is a phenomenon where a material exhibits spontaneous electric polarization that can be reversed by an external electric field, making these materials ideal for applications in memory storage and sensors. In layered materials, ferroelectricity arises from weak \hl{vdW} interactions between atomic layers, which enable long-range ordering and stable polarization  \cite{zhang2023ferroelectric}. \\
\noindent Poudel et al. (2019) explored the ferroelectric properties of black phosphorus and twelve monochalcogenide monolayers using eight different XC functionals, including traditional functionals as well as functionals with self-consistent vdW corrections, such as optPBE-vdW, optB86b-vdW, vdW-DF, vdW-DF2, and SCAN+rVV10. While the lack of experimental data limited the ability to identify the best XC functional, the benchmarking provided valuable insights. \hl{SCAN+rVV10 and optPBE-vdW delivered more favorable results, while LDA and PBE misrepresent lattice parameters and energy barriers} \cite{poudel2019group}. \\
\noindent Zhang et al. evaluated the performance of several XC functionals for ferroelectric perovskites, focusing on their ability to capture structural and electric properties. The study assessed LDA, PBE, hybrid functionals (HSE and B1-WC \cite{bilc2008hybrid}), and SCAN across ferroelectric and multiferroic materials with different bonding environments, providing a useful framework for judging functional transferability. \hl{Their results show that PBE tends to overestimate polarization and underestimate phonon frequencies, while LDA can underestimate polar distortions in hydrogen-bonded systems. In contrast, SCAN consistently improves structural properties, electronic polarizability, and ferroelectric distortions, indicating that meta-GGA functionals can better capture the coupled lattice-electronic response that controls ferroelectricity}. Although the benchmark focuses on perovskites, the same considerations are directly relevant to 2D ferroelectrics, where weak bonding, reduced dimensionality, and polar distortions make the choice of XC functional especially important \cite{zhang2017comparative}.
\subsubsection{Strategic approach}
Table \ref{table:structural_summary} \hl{highlights the functionals that consistently deliver the most accurate results for lattice constants, bond lengths, interlayer spacing, CDW, and ferroelectricity. These findings provide a strategic guide for selecting the most appropriate functional for various structural analyses in 2D materials.} \\
\begin{table}[h!]
\centering
\setlength{\arrayrulewidth}{0.35mm}
\renewcommand{\arraystretch}{1.25}
\caption{Best-performing XC functionals for major structural properties of 2D materials.}
\begin{tabular}{|p{5cm}|p{3.0cm}|p{2.8cm}|}
\hline \hline
\textbf{Structural Property} &
\textbf{Best Functional} &
\textbf{Second Best}
\\ \hline

Lattice constants & SCAN & r$^{2}$SCAN \\ \hline

Bond length & SCAN & PBEsol \\ \hline

Interlayer spacing (vdW-dominated) &
SCAN+rVV10 & optB86b-vdW \\ \hline

CDW-related lattice distortion & r$^{2}$SCAN & SCAN \\ \hline

Ferro-electricity & SCAN & SCAN+rVV10 
\\ \hline
\end{tabular}
\label{table:structural_summary}
\end{table}
\subsection{Electronic properties}
Predicting the electronic properties of 2D materials with experimental accuracy remains a significant challenge. While DFT provides a general picture of a material's electronic structure, traditional functionals like LDA and PBE often underestimate band gaps and misrepresent their nature. To improve these predictions, XC functionals must capture the unique quantum effects and enhanced electron interactions in 2D systems. Hybrid functionals typically offer more accurate descriptions of electronic properties, while many-body methods like the GW approximation provide a more precise treatment of QP energies. \hl{This section reviews the performance of various XC functionals in predicting the electronic properties of 2D materials, highlighting both their strengths and limitations.}
\subsubsection{Band Gap predictions}
\begin{figure}[h!]
\centering
    \includegraphics[scale=0.375]{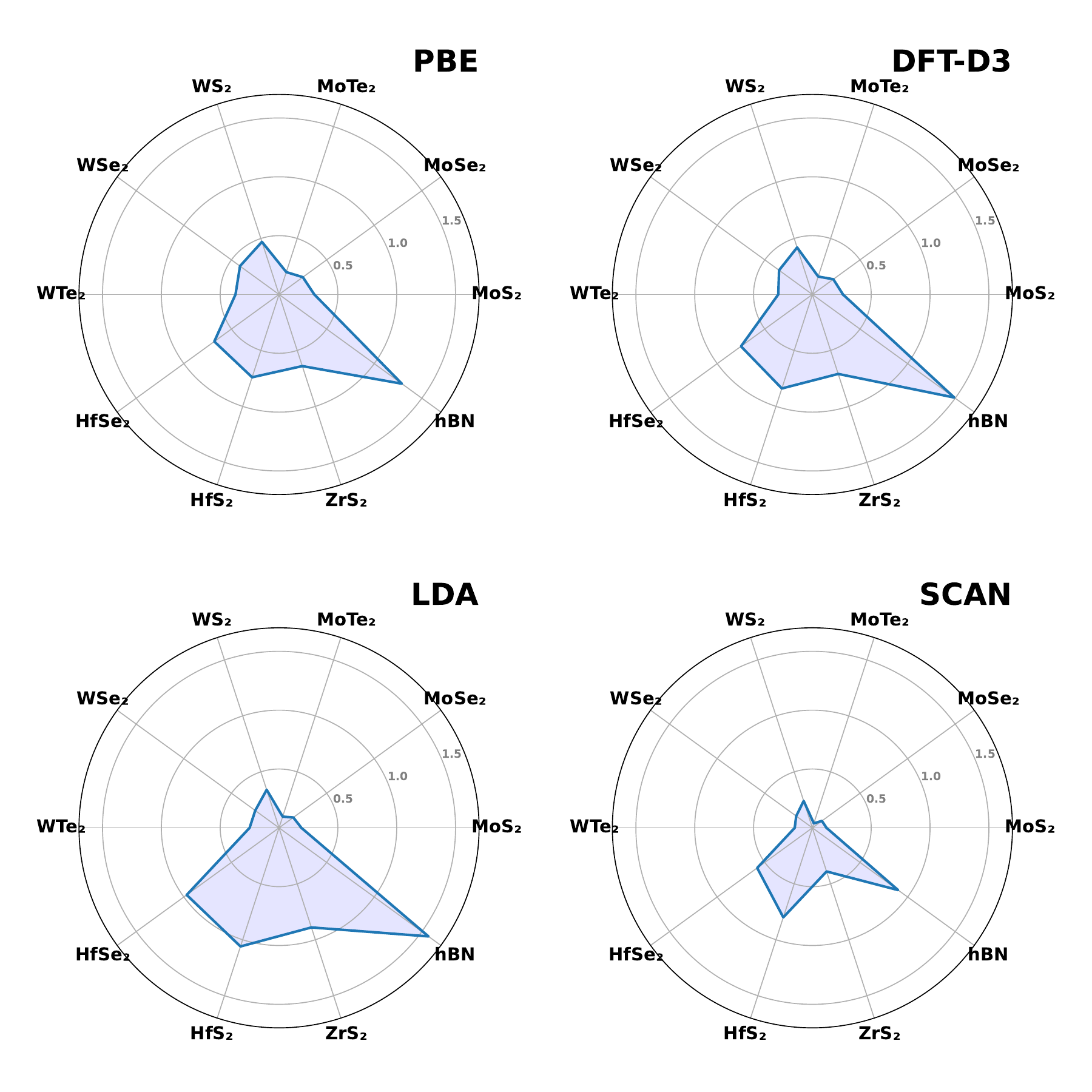}
    \caption{Radar plots showing mean absolute error (MAE) in band-gap predictions for several TMDC monolayers. Reprinted from \cite{javed2025structural}, copyright (2025), with permission from Elsevier.}
    \label{fig:radial}
\end{figure}

\begin{figure}[h!]
\centering
    \includegraphics[scale=0.5]{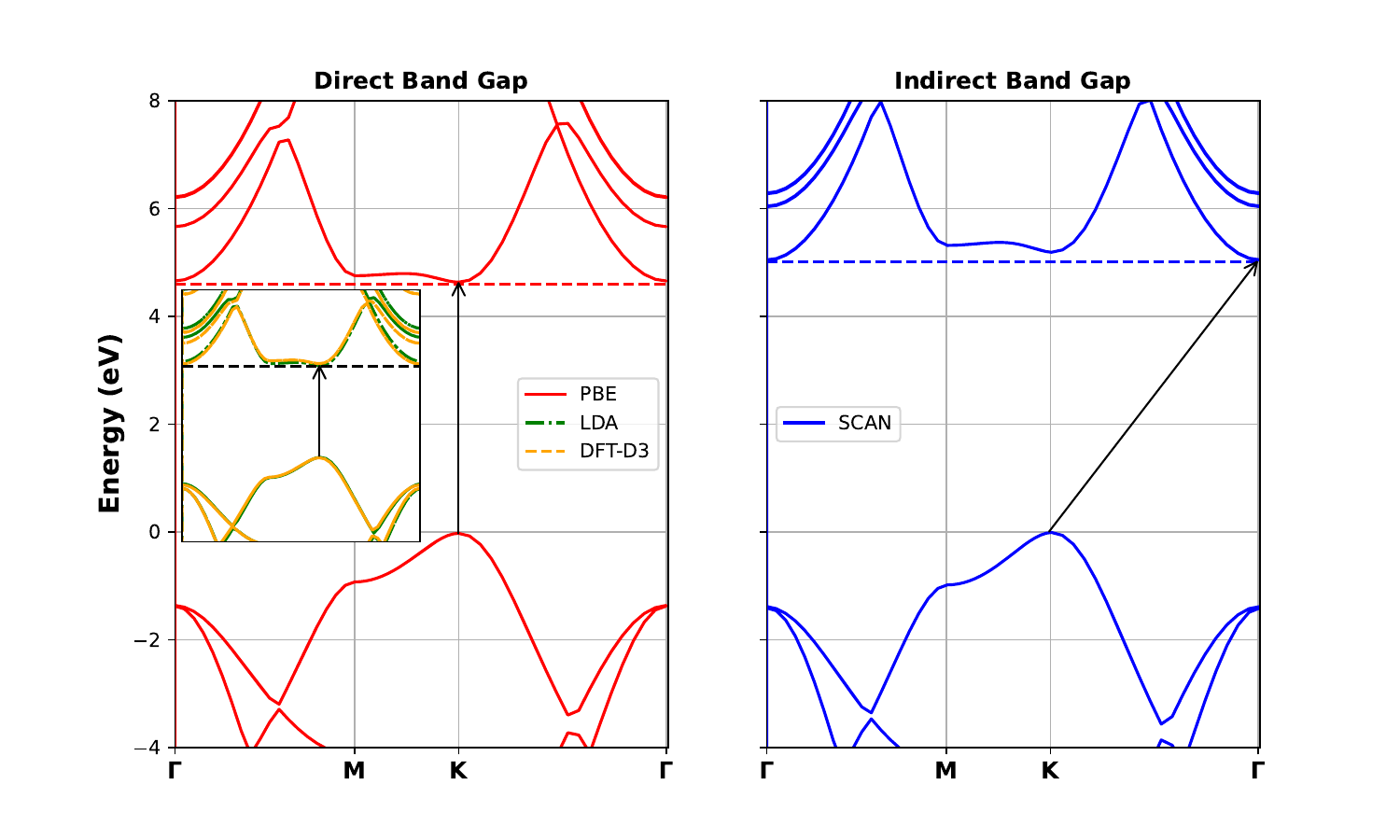}
    \caption{Band structure of hBN monolayer for different functionals. Reprinted from \cite{javed2025structural}, copyright (2025), with permission from Elsevier.}
    \label{fig:hbn}
\end{figure}

The band gap is a key electronic property that dictates whether a material behaves as a semiconductor or an insulator. It's important to note that \hl{experimental band gaps are optical in nature and differ from photoemission gaps due to the EBE, which can be computed using the Bethe–Salpeter equation.} \hl{EBE represents the energy required to separate a bound electron-hole pair into free charge carriers.} 
\begin{equation}
    E_{{electronic}} = E_{{optical}} + EBE
\end{equation}
\hl{However, for most materials, the excitonic energy is typically smaller than the errors introduced by DFT functionals, so we use experimental values as a reference for evaluating functional performance} \cite{radial_mae}. Table \ref{table:bandgaps_tmdcs} lists band gaps for pristine 2D materials, calculated using various functionals. As shown, semi-local functionals generally underestimate band gaps due to their lack of non-local exchange and their failure to capture large EBEs arising from reduced dielectric screening in these materials. The SCAN functional offers more accurate predictions than PBE and LDA, slightly underestimating experimental values by around 5\%, while reducing self-interaction errors, short-range vdW effects, and providing a better description of correlation effects. For further insight into the performance of these functionals, a statistical analysis of the band gap calculations is presented in Fig. \ref{fig:radial}, where the smallest area corresponds to SCAN, indicating its better performance relative to other functionals.
\begin{table}[h!]
\setlength{\arrayrulewidth}{0.35mm}
\setlength{\tabcolsep}{9.5pt}
\renewcommand{\arraystretch}{1}
\centering
\caption{\hl{Band gap (eV) of pristine monolayers, `I' indicates indirect band gap and `D' is for direct band gap. The band gap values not referenced indicates self calculated results and are in accordance with literature.}}
\label{table:bandgaps_tmdcs}
\begin{tabular}{ |c|c|c|c|c|c|c| }
\hline \hline
& \multicolumn{1}{|c|}{\textbf{LDA}} & \multicolumn{1}{|c|}{\textbf{PBE}} & \multicolumn{1}{|c|}{\textbf{SCAN}} & \multicolumn{1}{|c|}{\textbf{HSE06}} & \multicolumn{1}{|c|}{\textbf{GoWo}} & \multicolumn{1}{|c|}{\textbf{\hl{Experiment}}} \\
\hline 
\textbf{hBN} & 4.38 & 4.66 & 5.05 & 5.68 & 7.12 & 5.95 (I) \cite{fu2025indirect, hbn_indirect_2016} \\ \hline
\textbf{GaS} & -- & 2.50 \cite{jung2015red} & -- & 3.33 \cite{jung2015red} & 3.88 \cite{yagmurcukardes2016mechanical} & 3.05 (I) \cite{yagmurcukardes2016mechanical,gutierrez2021exploring} \\ \hline
\textbf{SnSe} & -- & 2.17 & 2.44 & 2.79 & 3.67 \cite{c2db_recent_2021} & 2.12 (I) \cite{yue2024identification} \\ \hline
\textbf{WS$_2$} & 1.67 & 1.58 & 1.75 & 2.06 & 2.53 & 1.99 - 2.02 (D) \cite{bg_ws2} \\ \hline
\textbf{HfS$_2$} & 0.90 & 1.02 & 1.16 & 2.13 & 2.94 & 1.96 (I) \cite{bg_HfS2} \\ \hline
\textbf{MoSi$_2$N$_4$} & 1.74 & 1.79 & 1.90 & 2.35 \cite{bafekry2021mosi2n4_theor.} & 3.17 \cite{liang2022highly_mosi2n4GW} & 1.94 (I) \cite{hong2020chemical_mosi2n4_Exp.} \\ \hline
\textbf{MoS$_2$} & 1.69 & 1.58 & 1.67 & 1.96 & 2.53 & 1.85 - 1.91 (D) \cite{eg_exp_mos2_2010} \\ \hline
\textbf{ZrS$_2$} & 0.91 & 1.16 & 1.41 & 2.17 & 2.89 & 1.80 (I) \cite{bg_zrs2} \\ \hline
\textbf{WSe$_2$} & 1.40 & 1.22 & 1.47 & 1.73 & 2.10 & 1.64 \cite{bg_wse2, bg_wse2_2} \\ \hline
\textbf{MoSe$_2$} & 1.42 & 1.32 & 1.47 & 1.80 & 2.12 & 1.55 - 1.59 (D) \cite{bg_mose2} \\ \hline
\textbf{MoTe$_2$} & 1.03 & 0.93 & 1.09 & 1.37 & 1.56 & 1.16 (D) \cite{bg_mote2_2014, bg_mote2_2015} \\ \hline
\textbf{HfSe$_2$} & 0.16 & 0.45 & 0.55 & 1.60 & 2.12 & 1.13 (I) \cite{bg_hfse2} \\ \hline
\textbf{WTe$_2$} & 0.85 & 0.73 & 0.95 & 1.14 & 1.38 & $\sim$ 1.00 (D) \cite{bg_wte2_mote2} \\ \hline
\textbf{NbS$_2$} & 0 & 0 & 0 & - & - & Metallic \cite{c2db_recent_2021} \\ \hline
\textbf{CNOH} & -- & 5.04 (D) & -- & 6.64 (D) \cite{zhang2025theoretical} & -- & -- \\ \hline
\textbf{SiNOH} & -- & 4.77 (D) & -- & 6.22 (D) & -- & -- \\ \hline
\textbf{PbNOCl} & -- & 0.16 (D) &  -- & 0.54 (D) & -- & -- \\ \hline
\textbf{SiNTeCl} & -- & 0.48 (I) &  1.54 (I) & 0.87 (I) \cite{zhang2025theoretical} & -- & -- \\ \hline
\end{tabular}
\end{table}
\noindent The hybrid functional HSE06 is well-regarded for its ability to accurately predict band gaps, especially for 2D materials. By incorporating a portion of exact exchange, HSE06 improves the electronic structure description compared to semi-local functionals, particularly in systems with strong SOC \cite{hybrid_2011, Becke_Hybrid:93}. \hl{While HSE06 provides accurate band gaps for many materials, it tends to slightly overestimate them. For example, it overestimates by $\sim$5\% for WS$_2$, $\sim$9\% for GeC and $\sim$20\% for MoSi$_2$N$_4$ monolayers} (Table \ref{table:bandgaps_tmdcs}). This overestimation can limit its effectiveness in systems with strong electron correlation or significant excitonic effects. \\
\noindent The G$_o$W$_o$ quasiparticle gap tends to overestimate the experimental band gap, \hl{because the weak screening in 2D materials leads to stronger electron‑hole interactions. These interactions create an attractive force between the quasi‑electron and quasi‑hole, which can significantly reduce the effective band gap. To accurately capture this effect, it is necessary to solve the BSE, which accounts for the excitonic effects by treating the attractive Coulomb interaction between the electron and hole} \cite{Ashwin_2012, EBE2017}. \\
\noindent The nature of the semi-conducting band gap, whether direct or indirect, is crucial, yet standard DFT functionals often fail to predict it accurately. \hl{For instance, the hBN monolayer, which exhibits an indirect band gap} \cite{hbn_indirect_2016}, \hl{is incorrectly predicted to have a direct band gap by PBE} \cite{hbn_direct}, \hl{LDA, and DFT-D3. In contrast, SCAN correctly identify the indirect band gap}, as shown in Fig. \ref{fig:hbn}.
\vspace{-0.25cm}
\newpage
\subsubsection{Spin orbit coupling and Valence band splitting} 
\begin{figure}[h!]
    \centering
    \includegraphics[scale=0.7]{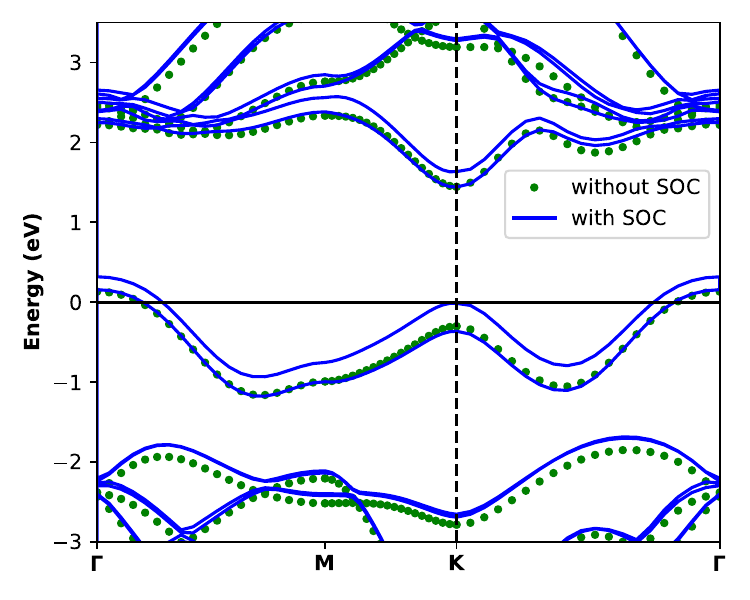}
    \caption{Comparative band structure of NbS$_2$ monolayer, with and without SOC.}
    \label{fig:soc_nonsoc}
\end{figure}

Spin-orbit coupling describes the interaction between an electron's spin and its orbital motion around the nucleus. This phenomenon becomes particularly prominent in materials composed of heavy atoms (e.g., W, Mo), where relativistic effects enhance the strength of SOC, paving the way for spintronic applications \cite{zollner2019strain}. In 2D materials, reduced dimensionality amplifies SOC effects \cite{chen2021spin}, causing significant spin splitting in the valence band. \hl{For example, the NbS$_2$ monolayer shown in Fig. \ref{fig:soc_nonsoc} is a metallic 2D material, and SOC lifts band degeneracies near the $K$ point and produces spin-dependent} \hl{band splitting around the Fermi level.} \hl{Another example is stanene, which is gapless without SOC, whereas SOC opens a gap of about 0.1 eV at the Dirac point, converting it into a quantum spin Hall insulating state} \cite{xu2013large_gap_stanene}. This is a critical feature for spintronic and valleytronic applications, where controlling spin and valley degrees of freedom is essential for developing next-generation electronic devices. \\
\noindent \hl{The level of theory significantly influences the sensitivity of valence band splitting. For WSe$_2$, the experimental VBS is approximately 520 $meV$} \cite{SOC_wse2}, \hl{while PBE predicts 430 $meV$, HSE06 estimates 580 $meV$, and G$_0$W$_0$ provides 540 $meV$. Experimental VBS values for 2D materials closely align with G$_0$W$_0$ calculations} \cite{javed2025investigation}.
\subsubsection{Band Gap Crossover}
Pristine monolayers from 2D families like TMDCs, SINOX, and MXenes are predominantly direct band-gap semiconductors \cite{2D_npj_overview_2020, 2d_tmdcs_2017}, with \hl{applications in optoelectronics and sensors, while indirect band-gap materials are key for thermo-electrics and photovoltaics}. Some 2D materials undergo a transition from an intrinsic direct to indirect under strain, revealing unique properties \cite{chiral_phonon_2024}. \\
\noindent Standard DFT functionals often fail to accurately predict the critical strain for this band-gap crossover. For instance, the MoS$_2$ monolayer transitions from a direct to an indirect band gap at uniaxial strain of $\sim$1.8\% \cite{evidence1blundo2020} or biaxial strain of $\sim$1.6\% \cite{bi_petHo2019moderate}. As shown in Table \ref{table:bgco}, HSE06 predicts a critical uniaxial strain of 1.6\% and 1.4\% for biaxial strain, closely matching the experimental values. In contrast, PBE and SCAN underestimate these experimental findings \cite{das2014microscopic, ahmad2014comparative, 1per_nguyen2016effect}.
\begin{table}[h!]
\setlength{\arrayrulewidth}{0.35mm}
\setlength{\tabcolsep}{9.5pt}
\renewcommand{\arraystretch}{1}
\centering
\caption{\hl{Equivalent strain (\%) for band cross-over ($\epsilon_\text{transition}$) in monolayer MoS$_2$.} Reproduced from \cite{javed2024band}, copyright (2024), with permission from Elsevier.}
\label{table:bgco}
\begin{tabular}{ |c|c|c|c|c|c|c| } 
 \hline \hline
\textbf{strain type} & \multicolumn{4}{c}{\textbf{critical strain (\%)}} & \multicolumn{2}{|c|}{\textbf{literature}} \\
 \hline
& PBE & SCAN & SCAN+rvv10 & HSE06 & Experimental & \hl{Theoretical} \\ \hline
uniaxial & 0.55 & 0.90 & 0.90 & 1.60 & 1.8 $\pm$ 0.7 & $<$\ 0.8  \\ \hline
biaxial & 0.30 & 0.60 & 0.60 & 1.40 & 1.6 $\pm$ 0.4 & $<$\ 0.9 \\ \hline
\end{tabular}
\label{table:strain}
\end{table}
\subsubsection{Is HSE06 a Holy Grail?}
\begin{figure}[h!]
    \centering
    \includegraphics[scale=0.55]{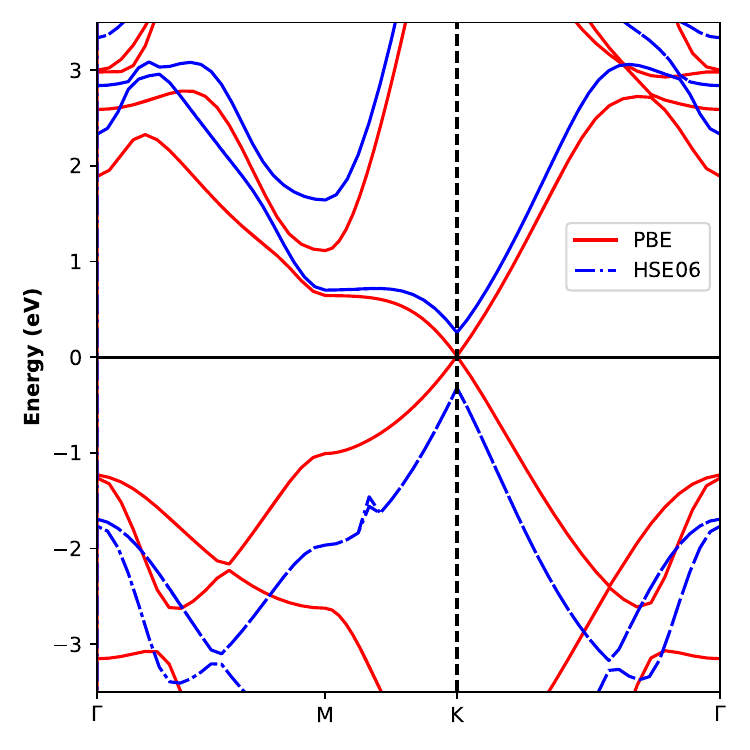}
    \caption{Comparative band structure of Silicene monolayer for PBE and HSE06 functional. Clearly HSE06 hybrid functional results in artificial band gap opening.}
    \label{fig:gapless}
\end{figure}

Despite the widespread use of HSE06 for band gap predictions of 2D materials, it has notable limitations. For instance, when applied to gapless 2D materials like \hl{graphene and silicene, HSE06 often artificially opens a small band gap (0.1–0.3 eV) in materials that are experimentally gapless} \cite{maroof2025first, ni2012tunable}, distorting the linear Dirac cone at the K-point. While this behavior aligns with the non-local HF exchange component of the functional, it contradicts experimental observations and underscores a key limitation of hybrid functionals that they can misrepresent delocalized $\pi$-electron systems, as shown in Fig. \ref{fig:gapless}.
\newpage
\noindent \hl{In the case of FeCl$_2$, HSE06 predicts a band gap of 3.26 $eV$, significantly larger than the experimental value of 1.2 eV} \cite{lu2022unique}. Although it improves on PBE, which predicts a much smaller gap of 0.35 $eV$, it still does not match the experimental value. In contrast, the SCAN functional predicts a band gap of 0.77 $eV$, closer to the experimental value. The LMBJ potential, though not fully self-consistent, provides the most accurate prediction at 1.38 $eV$, outperforming HSE06. This trend is consistent across all FeX$_2$ monolayers (X = Cl, Br, I). \\
\noindent Thus, while HSE06 is a powerful tool, it is not a one-size-fits-all solution and should be applied carefully, especially in systems where symmetry-protected electronic states are critical.
\subsubsection{Strategic approach}
\hl{This section evaluates the best-performing XC functionals for key electronic properties of 2D materials. As shown in Table} \ref{table:electronic_summary}, \hl{HSE06 excels in predicting band gap magnitude, nature, and crossover, with SCAN providing a close alternative at a significantly lower computational cost. G$_0$W$_0$ offers the most accurate results for incorporating excitonic and SOC effects. For gapless materials like graphene and silicene, PBE and LDA show reliable performance.}
\begin{table}[h!]
\centering
\setlength{\arrayrulewidth}{0.35mm}
\renewcommand{\arraystretch}{1.25}
\caption{Best-performing XC functionals for major electronic properties of 2D materials.}
\begin{tabular}{|p{5.4cm}|p{3.0cm}|p{2.8cm}|}
\hline \hline 
\textbf{Electronic Property} &
\textbf{Best Functional} &
\textbf{Second Best}
\\ \hline
Band gap magnitude, nature (direct vs indirect) and cross-over (direct to indirect) &
HSE06 & SCAN \\ \hline

Quasiparticle band gap & G$_0$W$_0$ &
LMBJ / mTASK \\ \hline

Valence band splitting &
G$_0$W$_0$+SOC & PBE+SOC \\ \hline

Gapless materials (graphene, silicene) &
PBE & LDA \\ \hline
\end{tabular}
\label{table:electronic_summary}
\end{table}
\subsection{Optical properties}
Two-dimensional materials exhibit optical properties that are profoundly different from their bulk counterparts due to strongly bound excitons and large optical gaps \cite{xia2014two, bernardi2017optical}. As reduced dielectric screening and quantum confinement amplify Coulomb interactions between electrons and holes, standard density‐functional approximations severely underestimate optical gaps and exciton binding energies. To better understand the limitations of these methods, we compare different approaches for calculating optical properties in 2D materials. 
\begin{itemize}
    \item \textbf{Independent‐particle calculations:} Semilocal functionals such as LDA and GGA underestimate band gaps and neglect electron–hole interactions, making them incapable of capturing excitonic effects. For instance, DFT-PBE predict an indirect band gap of $4.66~eV$ for monolayer hBN, whereas the experimental optical gap is approximately $6~eV$ (Table \ref{table:bandgaps_tmdcs}) with an exciton binding energy of around $2~eV$ \cite{galvani2016excitons}.
    \item \textbf{Hybrid functionals:} Hybrid functionals partially correct the band-gap underestimation of semilocal functionals but still fail to capture excitonic effects. For some 2D monolayers (Table \ref{table:bandgaps_tmdcs}), the band gaps obtained from hybrid functional calculations are close to the quasiparticle values; however, the corresponding exciton binding energies remain significantly smaller than those predicted by BSE calculations.
    \item \textbf{GW+BSE framework: } The GW approximation corrects quasiparticle band gaps by incorporating the self-energy of electrons, which is crucial for obtaining accurate band structures. The BSE method solves the electron–hole interaction by describing the two-particle Green's function and capturing the correlation between the electron and hole, and it provides a more accurate description of optical properties. When combined, the GW and BSE form a robust theoretical framework for predicting the optical behavior of 2D materials, explicitly accounting for excitonic effects, optical absorption, and other many-body phenomena. For most 2D materials, the BSE optical gap is $1–2~ eV$ lower than the GW band gap, making BSE the method of choice for properly accounting for electron-hole interactions and yielding accurate optical spectra and exciton binding energies.
\end{itemize}
\subsubsection{Early GW+BSE studies}
\begin{figure}[h!]
\centering
    \includegraphics[scale=0.325]{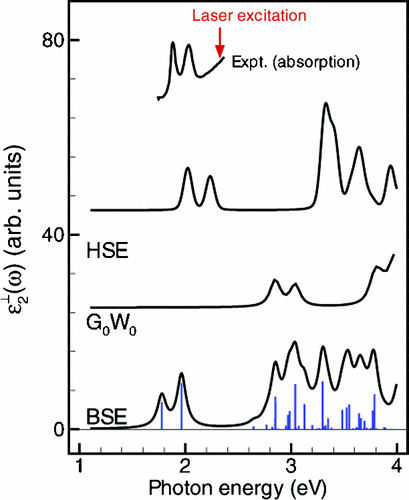}
    \caption{Imaginary part of transverse dielectric constant for monolayer MoS$_2$, as a function of photon energy. Computed spectra are presented for three levels of theory: HSE06 functional, G$_0$W$_0$, and BSE. Vertical (blue) bars represent the relative oscillator strengths for the optical transitions. The two lowest-energy peaks in the spectrum (first two bars) correspond to the A and B excitons. As seen, the closest agreement with experiments is obtained at the level of BSE calculations. Reprinted from \cite{Ashwin_2012}, copyright (2012), with permission from American Physical Society.}
    \label{fig:ashwin}
\end{figure}

\begin{figure}[h!]
    \centering
    \includegraphics[scale=0.5]{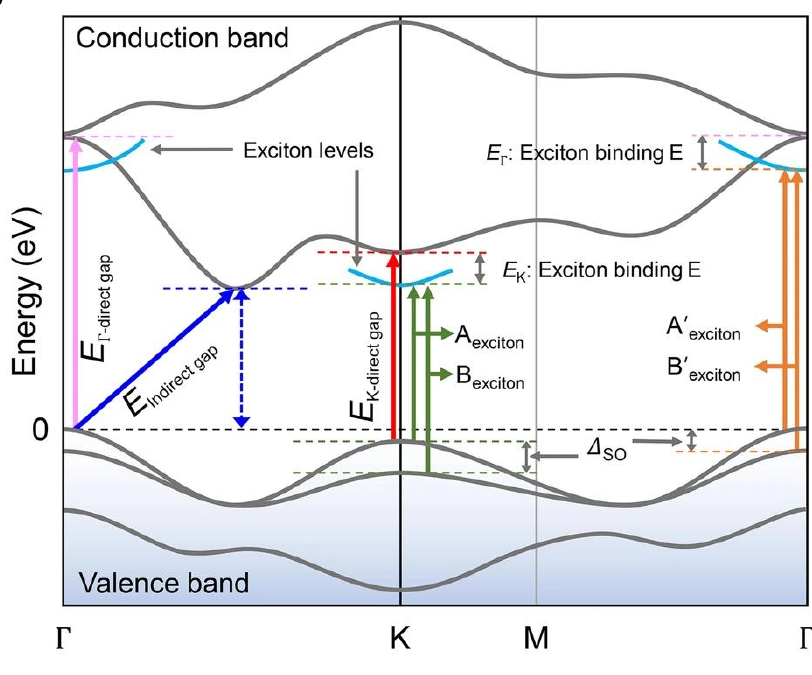}
    \caption{Schematic representation of the band structure in multilayer TMDCs with corresponding optical transitions labeled. Reprinted from \cite{jung2022unusually}. \href{https://creativecommons.org/licenses/by/4.0/}{\underline{CC BY 4.0}}.}
    \label{fig:enter-ebe}
\end{figure}

\noindent One of the earliest works on the GW+BSE framework by R. Ashwin (2012) \cite{Ashwin_2012} investigates the significant excitonic effects in monolayers of molybdenum and tungsten dichalcogenides. \hl{The study emphasizes the importance of incorporating many-body effects to accurately capture the true nature of the A and B excitons in 2D materials. The experimental absorption spectrum for MoS$_2$ in this study shows the closest agreement with experimental results when calculated using the BSE approach}, as shown in the Fig. \ref{fig:ashwin}. The $A$ exciton corresponds to the lowest-energy transition at the $K$ point, while the $B$ exciton arises from transitions at the $K'$ point as shown in Fig. \ref{fig:enter-ebe}. These excitons play a crucial role in the optical properties of 2D materials. This figure also emphasizes the significant spin-orbit coupling in these materials, which causes distinct band splitting. Exciton peaks i.e., $A$ and $B$ near the $K$-point and $A'$ and $B'$ near the $\Gamma$-point are shown with symmetric absorption bands, in contrast to the asymmetric absorption associated with these band gaps.
\subsubsection{Recent advances in GW+BSE studies}
\begin{figure}[h!]
    \centering
    \includegraphics[scale=0.23]{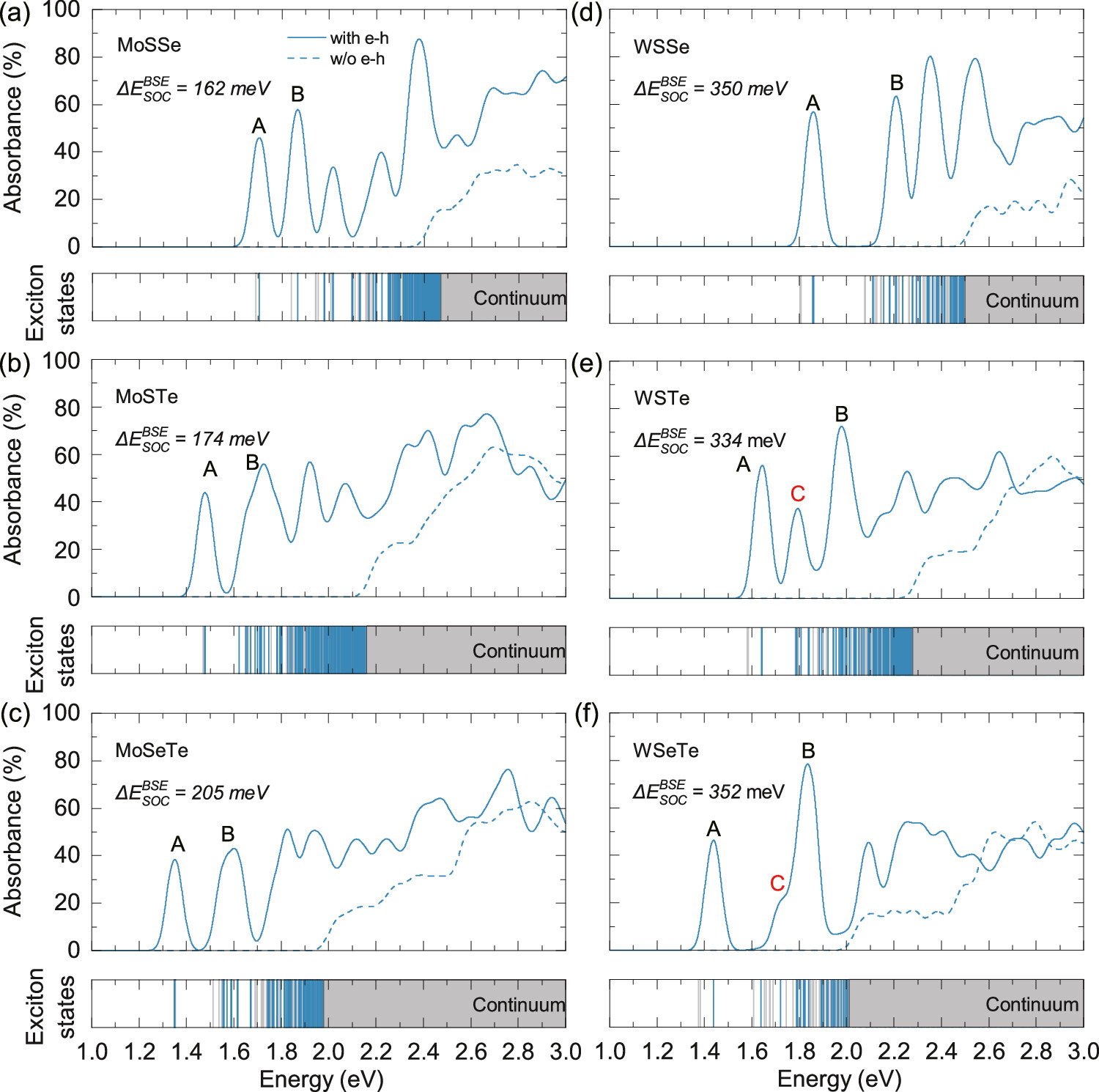}
    \caption{Optical absorbance with (blue solid line) and without (blue dashed line) electron–hole interactions and exciton spectrum of Janus MoSSe (a), MoSTe (b), MoSeTe (c), WSSe (d), WSTe (e), and (f) WSeTe. The blue vertical lines and gray vertical lines in the exciton spectrum are for the bright and dark exciton states, respectively. Reprinted from \cite{hu2025quasiparticle}. \href{https://creativecommons.org/licenses/by/4.0/}{\underline{CC BY 4.0}}.}
    \label{fig:janus}
\end{figure}

\noindent In a recent study, Qian et al. (2025) compare the EBE of regular TMDCs with their Janus counterpart using the GW+BSE framework. Their findings indicate that Janus monolayers of these materials exhibit significantly higher EBE compared to typical TMDCs. Specifically, \hl{the absorption edges in MoSSe and WSSe are observed at 1.70 $eV$ and 1.87 $eV$ respectively, aligning well with experimental photoluminescence (PL) values of 1.68 $eV$ and 1.87 $eV$.} In addition, the study reports notable splitting of $A$ and $B$ excitonic peaks, with a newly identified $C$ peak in the W series as shown in Fig. \ref{fig:janus}, which suggests the potential of Janus TMDCs for valleytronic applications \cite{hu2025quasiparticle}.
  \\
\noindent \hl{C$_2$N, a nitrogenated two-dimensional carbon material, exhibits pronounced thickness-dependent excitonic effects. LDA predicts a band gap of about 1.7 $eV$, which is significantly underestimated due to the neglect of nonlocal electron–electron interactions. In contrast, GW+BSE calculations yield a QP gap of 3.75 $eV$ and a first bright exciton at 2.64 $eV$, corresponding to a binding energy of approximately 1.11 $eV$. As additional layers are added, enhanced dielectric screening progressively reduces the EBE to 0.76 $eV$ in the bilayer and 0.53 $eV$ in the trilayer. In the bulk limit, as dielectric screening further strengthens with increasing thickness, the optical gap decreases to 2.02 $eV$ with a negligible binding energy of 0.04 $eV$, closely matching the experimental gap (1.96 eV)} \cite{sun2016many}. \\
\noindent Tang et al. (2025) combine advanced meta-GGA functionals, SCAN and mTASK, with a dielectric function-based BSE (mBSE) method to predict the optical response properties of bulk and monolayer 2D materials \cite{tang2025meta}. This framework effectively captures the screening effect in materials while maintaining computational efficiency. Their findings demonstrate the accuracy of this method in predicting optical absorption spectra for MoS$_2$ and hBN monolayers, with results closely matching experimental data.
\subsubsection{Strategic approach}
\noindent \hl{The GW+BSE approach serves as a state-of-the-art method for accurately describing the optical properties of two-dimensional materials, as it accounts for both quasiparticle corrections to the electronic band structure through the GW approximation and excitonic effects via the BSE.}
\subsection{Magnetic properties}
\begin{figure}[h!]
    \centering
    \includegraphics[scale=0.48]{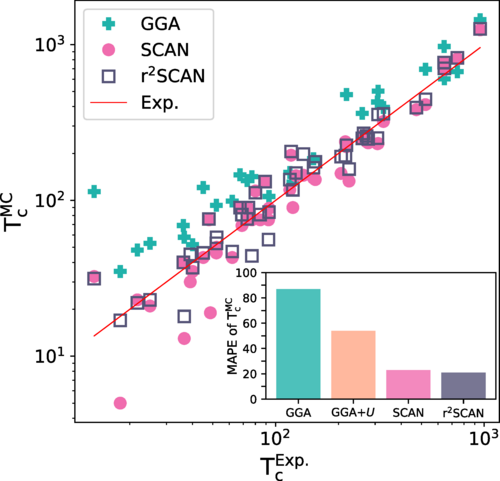}
    \caption{A comparison of the computed Néel transition temperatures using the SCAN and r$^2$SCAN functionals against experimental transition temperatures. The inset figure shows the total mean absolute percentage error (MAPE) of transition temperatures for different functionals. Reprinted from \cite{rezaei2025evaluating}, copyright (2025), with permission from American Physical Society.}
    \label{fig:mag}
\end{figure}

Magnetic properties in 2D materials are crucial for spintronic devices, offering breakthroughs in energy-efficient electronics and quantum technologies. However, the Mermin–Wagner theorem \cite{mermin1966absence} predicts that continuous symmetry cannot support long-range order in 2D Heisenberg models. This limitation is overcome by anisotropy, which stabilizes magnetic ordering in low-dimensional systems even at finite temperatures \cite{aldea2025_magnetism_review}. This was first demonstrated by the discovery of the 2D magnetic monolayer CrI$_3$ in 2017, with a Curie temperature of 45 K \cite{lado2017origin}. The later discoveries of ferromagnetism in bilayer Cr$_2$Ge$_2$Te$_6$ \cite{gong2017discovery} and monolayer Fe$_3$GeTe$_2$ \cite{fei2018two} further confirm stable ferromagnetism in ultrathin materials.  \\ 
\noindent A recent study (2025) benchmarked the SCAN family against GGA, GGA+U, and HSE06 for predicting transition temperatures in 48 antiferromagnetic bulk and 2D materials \cite{rezaei2025evaluating}. \hl{Transition temperatures computed from exchange constants obtained with SCAN and r$^2$SCAN showed Pearson correlation coefficients of 0.97 and 0.98 with experimental values, significantly outperforming GGA and GGA+U. In contrast, HSE06 underestimated transition temperatures due to its screening of exchange interactions}. As shown in Fig. \ref{fig:mag}, SCAN and r$^2$SCAN consistently outperform other functionals in predicting transition temperatures. However, discrepancies remain; for example, SCAN correctly identifies CrF$_2$ as antiferromagnetic, while r$^2$SCAN incorrectly predicts it as ferromagnetic. Similarly, the transition temperatures for MnTe differ notably between these functionals.
 \\
\noindent Tellez‑Mora et al. (2024) \cite{tellez2024systematic} introduces a systematic workflow that combines DFT with linear spin-wave theory (LSWT) to predict stable magnetic configurations. This involves calculating the exchange coupling tensors using the Green`s function method (using TB2J code) and applying them in a self-consistent loop to minimize the spin-wave eigenvalues and thus optimizing the magnetic structure. \hl{The workflow was tested on a variety of materials, including NiO (an antiferromagnet), FePS$_3$ and FeCl$_2$ (a 2D layered vdW material), as well as MnF$_2$, FeP, and CuO. The method provides reliable predictions of the magnetic ground states, showing strong agreement with experimental results.} This approach enables to predict magnetic configurations with greater accuracy than traditional heuristic methods. Despite the advancement of this methods, a key difficulty is incorporating the effects of single-ion anisotropy in low-dimensional systems, which are crucial for capturing the full range of magnetic behaviors. \\
\noindent A 2020 study applied the Quasiparticle self-consistent GW (QSGW) method to magnetic 2D vdW materials i.e., VI$_3$, CrI$_3$, CrGeTe$_3$ and Fe$_3$GeTe \cite{lee2020role}. The QSGW method treats the XC energy as a non-local self-energy, avoiding the Hubbard-U parameter used in DFT+U calculations. \hl{While conventional DFT+U could not simultaneously reproduce the electronic band structures and magnetic properties of these 2D magnets, QSGW provided band structures, DOS, and spin splittings that were consistent with available experimental data. Additionally, QSGW accurately predicted the semiconducting states of VI$_3$ whereas both DFT and G$_0$W$_0$ failed to do so.} However, QSGW is computationally expensive and still requires careful treatment of SOC to capture magnetic anisotropy. 
\noindent Thus, for magnetic properties, meta-GGAs offer great potential in terms of both computational cost and accuracy, with significant room for further exploration in this direction.
\subsection{Thermal properties}
\begin{figure}[h!]
    \centering
    \includegraphics[scale=0.33]{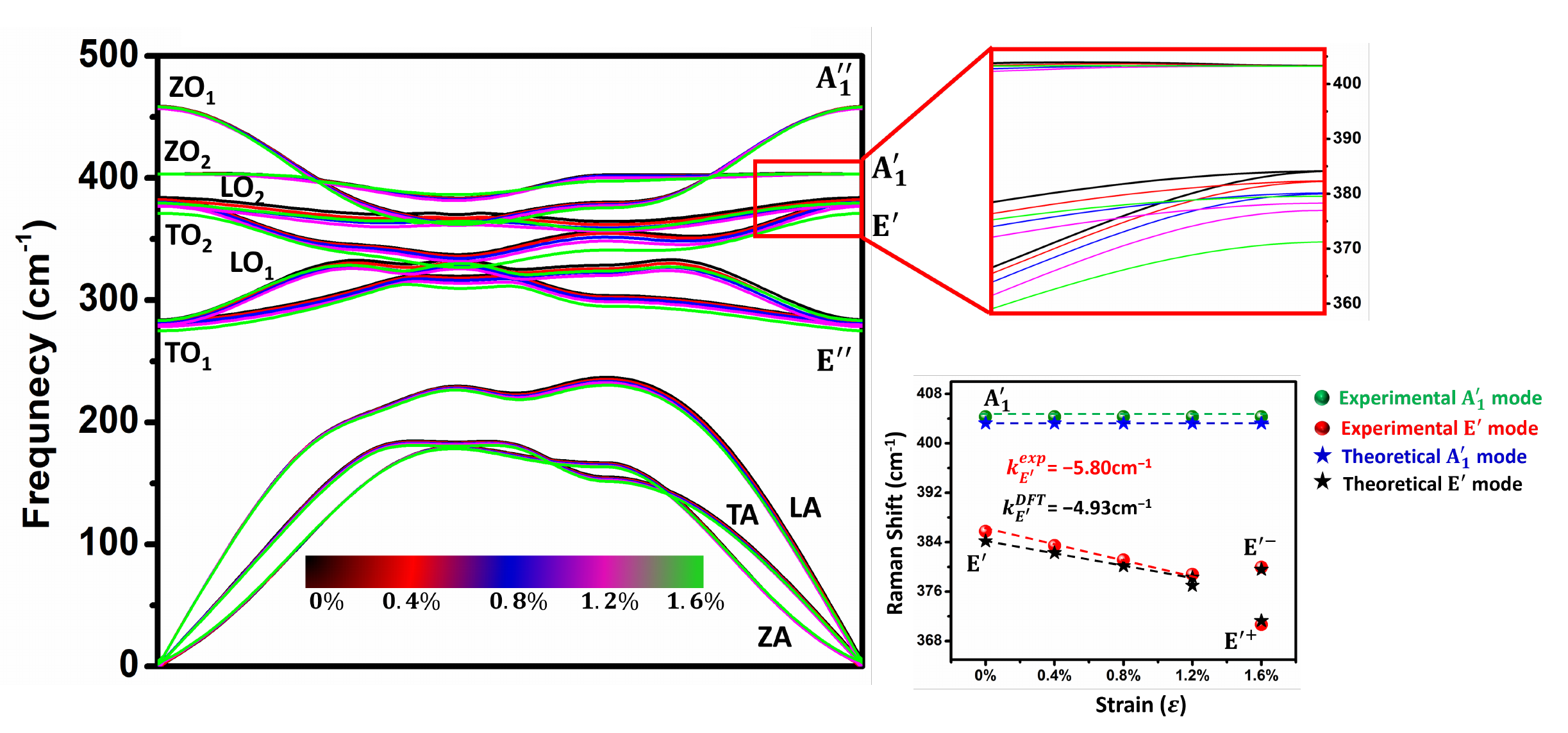}
    \caption{Phonon dispersion of monolayer MoS$_2$ under 0\%-1.6\% strain shows redshift in E$^{'}$/E$^1_{2g}$ modes, with no change in monolayer A$_1^{'}$ mode. Reproduced from \cite{ali2024room}, Copyright (2024), with permission from AIP Publishing.}
    \label{fig:apl}
\end{figure}

The thermal properties of 2D materials play a critical role in their potential applications in energy conversion and heat management. They exhibit unique behaviors compared to their bulk counterparts, including enhanced thermal conductivity ($\kappa$) and low thermal resistance. Graphene exhibits extraordinary lattice thermal conductivities exceeding $3000\  W m^{-1} K^{-1}$ at room temperature because of its strong covalent bonding and long phonon mean‑free paths \cite{renteria2014graphene}. \hl{On the other hand, 2D semiconductors such as MoS$_2$, WS$_2$ and black phosphorus have thermal conductivities below that of bulk silicon} ($\sim$ 142\ $W m^{-1} K^{-1}$) \hl{due to their heavy atoms and weak bonding which results in strong phonon scattering} \cite{gu2016layer, farooq2025phonons}. \hl{But the discovery of new 2D materials, such as MoSi$_2$N$_4$ from MA$_2$Z$_4$ family, has shown that this material exhibit high mobility and large thermal conductivity} ($\sim$ 173\ $W m^{-1} K^{-1}$ \cite{he2024unusually}). This demonstrates that 2D materials can exhibit thermal performance comparable to or even exceeding that of bulk materials. Thermoelectric applications require low value of $\kappa$, while heat spreaders demand high $\kappa$. Accurate theoretical predictions are essential, and the choice of the best XC functional within DFT is crucial. \\
\noindent In a recent study (2024), Akhter et al. \cite{ali2024room} used room-temperature Raman and photoluminescence spectroscopy to investigate the effects of uniaxial strain on monolayer MoS$_2$. The study found notable shifts and splittings in the E$_{2g}$ and A$_{1g}$ Raman peaks under strain. \hl{These experimental results were validated through DFT calculations using the PBE functional for both strained and pristine MoS$_2$ monolayer (Fig. \ref{fig:apl}), showing good agreement.}
 \\
\noindent Phosphorene monolayer exhibits strong anisotropy which makes prediction of $\kappa$ extremely sensitive to XC functionals. Zhang et al. \cite{zhang2022anisotropy} performed first‑principles calculations for benchmarking using 12 different XC functionals (LDA, PBE, rPBE, PBEsol, PW91, two ultrasoft pseudopotentials, three optB functionals and two vdW‑DF functionals). All functionals predict larger $\kappa$ in the zigzag direction than in the armchair direction. LDA overbinds, giving stiffer phonons and higher $\kappa$, whereas PBE and PBEsol yield lower but more realistic values. The USPP based functionals gives very low thermal conductivities.
\begin{table}[h!]
\centering
\caption{Thermal conductivity of phosphorene computed using different XC functionals. Reproduced from \cite{zhang2022anisotropy}. \href{https://creativecommons.org/licenses/by/3.0/}{\underline{CC BY 3.0}}.}
\begin{tabular}{|l|c|c|}
\hline \hline
\textbf{Functional} & \textbf{$\kappa_{\text{zigzag}}$ (W/mK)} & \textbf{$\kappa_{\text{armchair}}$ (W/mK)} \\
\hline \hline
LDA        & 4.93  & 1.02 \\ \hline
PBE        & 14.89 & 2.42 \\ \hline
rPBE       & 11.81 & 2.31 \\ \hline
PBEsol     & 5.91  & 1.06 \\ \hline
PW91       & 7.22  & 1.62 \\ \hline
\hl{USPP-GGA}  & 0.51  & 0.15 \\ \hline
\hl{USPP-LDA}  & 5.81  & 1.59 \\ \hline
optB88     & 30.48 & 5.11 \\ \hline
optB86b    & 15.93 & 3.26 \\ \hline
optPBE     & 23.19 & 5.24 \\ \hline
\hl{vdW-DF2}   & 14.56 & 2.41 \\ \hline 
\end{tabular}
\end{table} \\
\noindent In lattice-dynamics calculations, the second-, third-, and fourth-order interatomic force constants (IFCs) describe increasingly complex interactions between atoms. The 2nd-order IFCs determine harmonic phonon frequencies, while the 3rd-order IFCs govern anharmonic phonon–phonon scattering, which mainly controls thermal conductivity. The 4th-order IFCs represent even higher-order anharmonicity and become important for accurately predicting phonon lifetimes and thermal transport, especially in materials with strong anharmonic effects \cite{li2023first, kielar2024anomalous, han2022fourphonon}. \\ 
\noindent The study by Zhou et al. demonstrates that, unlike the 2nd- and 3rd-order IFCs, the 4th-order IFCs are highly sensitive to the energy-surface roughness associated with different XC functionals \cite{zhou2023extreme}. This sensitivity varies across materials and can introduce large errors in the predicted $\kappa$. As a result, when computing 4th-order IFCs using finite-difference methods, the atomic displacement must be sufficiently large to overcome this roughness and yield reliable phonon lifetimes. For example, with LDA, PBE, and PBEsol, silicon behaves smoothly for lower-order IFCs but shows significant roughness in the 4th-order terms, leading to an underestimation of $\kappa$. In Boron Arsenide, only PBEsol exhibits roughness at the 4th-order level, causing a 40\% reduction in the predicted thermal conductivity. For NaCl, all three functionals produce rough 4th-order IFCs, resulting in a 70\% underprediction of $\kappa$ at room temperature \cite{zhou2023extreme}. \\
\noindent Recently, conventional (phonon-mediated) superconductivity in transition metal oxides, carbides, and nitrides, Shahbaz and coworkers highlighted the critical role of exchange-correlation functionals~\cite{bakar2023effects, Azam:24, Rehman_2024, Bakar:PdCPtC25}. \hl{They found that including nonlocal correlations accurately is essential, as it influences both the number of electrons available for Cooper-pair formation and the phonons that mediate their interaction. Consequently, accurate functionals not only correct electron-electron interactions but also provide a more reliable potential energy surface for the ion cores.} This leads to improved structural predictions and phonon frequencies that align more closely with experimental results. vdW-DF3 turned out to be a promising functional for accurate structures and phonon spectra.| There are a lot of 2D materials with a significant contribution from vdW interactions. Thus, nonlocal correlations would not only improve the electronic energy levels but also the structure and dynamical properties of these materials. The effects of more accurate meta-GGAs and hybrid functionals are unexplored, representing an open direction for future research.
\subsubsection{Strategic approach}
\noindent \hl{For thermal properties of 2D materials, PBE and PBEsol functionals generally offer reliable predictions, though further exploration is needed, as no comprehensive studies on meta-GGAs for thermal properties have been reported. The strong sensitivity of 4th-order IFCs to the choice of XC functional highlights the need for advanced methods to accurately capture phonon lifetimes and lattice thermal conductivity in low-dimensional systems.}
\newpage

\section{Machine learning approaches for enhancing theoretical predictions}
\begin{figure}[h!]
    \centering
    \includegraphics[scale=0.35]{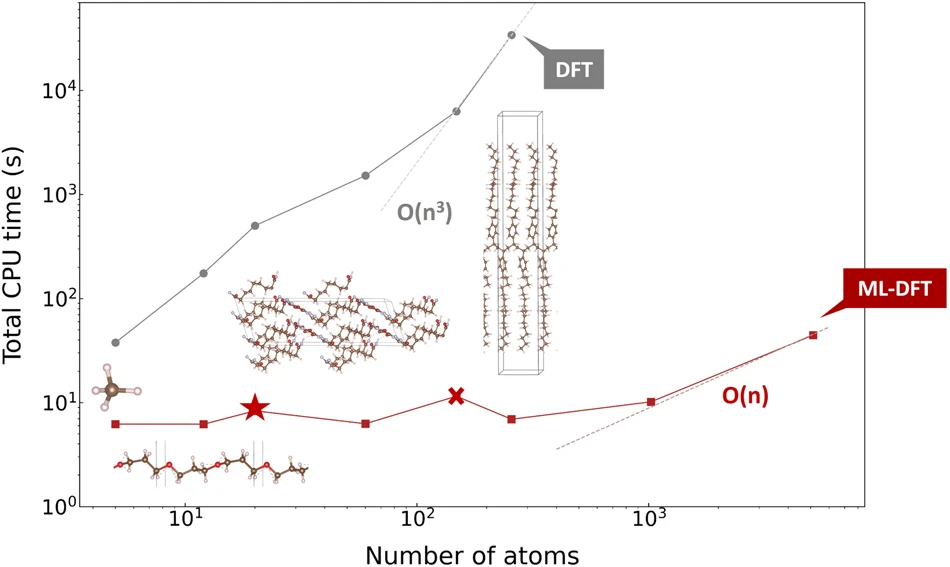}
    \caption{DFT shows an initial high computational cost and a cubic dependence on the system size. On the other hand, ML-DFT is orders of magnitude faster than DFT and linearly dependent on the system size. Red squares: structures with carbon and hydrogen. Red star: structure with three elements. Red cross: structure with all four elements. Reprinted from \cite{del2023deep}. \href{https://creativecommons.org/licenses/by/4.0/}{\underline{CC BY 4.0}}.}
    \label{fig:ml_cost}
\end{figure}

DFT remains a powerful and widely used method for predicting materials properties, but its high computational cost can limit studies of large, complex, or high-throughput 2D systems (Fig. \ref{fig:ml_cost}). \hl{Machine learning (ML) offers a complementary route by using data-driven algorithms to learn structure--property relationships from existing calculations and experiments. In materials informatics, these trained models can rapidly estimate structural, electronic, and optical properties, helping to screen candidate materials before more expensive DFT or many-body calculations are performed.}
\noindent Thus, ML enhances theoretical prediction by improving speed, scalability, and materials discovery while still relying on reliable reference data. In the context of XC functionals, ML is particularly useful because it can learn corrections to approximate functionals, improve XC potentials, and identify patterns in high-level reference data that are difficult to encode analytically. The following section reviews key contributions of machine learning in the prediction of layered material properties and in the improvement of XC-related theoretical methods \cite{lu2024machine, sajid2022spin}. 
 \\

\subsection{Transferability of ML models and XC corrections}
\noindent \hl{Transferability is one of the central challenges in applying ML to XC functional development. A model trained on a limited set of structures may perform well within that chemical space but fail for new bonding environments, dimensionalities, or defect configurations. This issue is especially important for 2D materials, where covalent bonding, vdW interactions, anisotropic screening, spin-orbit coupling, and reduced dimensionality can appear in the same material family. Therefore, transferable ML models should not only fit numerical data but also preserve physically meaningful trends across different materials classes.}

\noindent \hl{Recent progress in ML has addressed specific deficiencies of traditional functionals by learning corrections that respect known physical constraints. For example, Pokharel et al. propose training neural networks to reproduce the SCAN energy density using only density and its gradients, while enforcing uniform density-scaling, spin-scaling, and Lieb-Oxford bounds} \cite{pokharel2022exact}. \hl{These constraints ensure that the resulting ML functional remains physically consistent and improves transferability across chemical environments. The authors also highlight that poor performance in under-represented bonding regions is often due to inadequate training data, and that adding data or enforcing additional constraints can mitigate this issue. These findings illustrate that ML corrections can address transferability issues by embedding exact constraints and augmenting training sets with representative bonding motifs.} \\

\noindent Brockherde et al.’s \cite{dick2020machine} NeuralXC framework demonstrates that ML-corrected functionals can be highly data-efficient as they achieve MAE below 0.01 $eV$ with roughly 100 training samples, and the error saturates as more data are added. NeuralXC uses a neural network to learn corrections to PBE from the electron density and its derivatives. For example, the model trained on a small set of ethane and propane structures accurately predicts the energies of n-butane and isobutane, and adding a few relevant structures further reduces systematic errors, indicating promising configurational and limited compositional transferability. \\

\noindent Structure-aware deep transfer learning, developed by Gupta et al. (2024) using the ALIGNN graph neural network, improves predictive accuracy across diverse material datasets. The model first learns structural and chemical correlations from a large dataset. It is then fine-tuned on smaller, specialized datasets, enabling efficient prediction of properties such as band gaps, bulk and shear moduli, and elastic constants. By embedding atomic connectivity and local geometric information into the latent space, the network captures transferable physical relationships governing material behavior. This framework outperforms conventional models trained from scratch. Pre-training on general structural patterns ensures more robust predictions. Evaluations across 115 datasets showed that transfer learning models outperformed models trained from scratch in 90\% of cases. This work highlights the growing role of transfer learning in overcoming data scarcity while preserving the physics of structure--property relationships \cite{gupta2024structure}. \\

\subsection{Data requirements and reference-quality datasets}
\noindent \hl{The predictive power of ML depends strongly on the quality and diversity of the data used for training. For 2D materials, this is a key bottleneck because accurate reference data may require hybrid functionals, GW, BSE, quantum Monte Carlo, or carefully benchmarked experimental measurements. Standard PBE-level datasets are useful for rapid screening, but they may also transfer the limitations of the underlying functional into the ML model. Therefore, the construction of high-fidelity datasets is essential for training ML models that can improve, rather than simply reproduce, existing XC functional errors} \cite{javed2025structural}.

\noindent \hl{Descriptor-based ML has already shown how relatively compact datasets can be used efficiently for 2D materials. A 2023 study} \cite{dau2023descriptor} \hl{created vectorized descriptors from property matrices and empirical features, such as electronegativity, and trained ensemble methods like extreme gradient boosting using data from the C2DB. These models achieved coefficients of determination above 0.9 and mean absolute errors below 0.23 $eV$ for band-gap and work-function predictions. By combining database-derived features with simple descriptors, the authors minimized overfitting and demonstrated that small datasets can still produce accurate ML models.}

\noindent In a recent study, Javed et al. (2025) introduced an ML-based approach to predict EBEs in 2D materials, providing a computationally efficient alternative to traditional methods like many-body perturbation theory (GW) and the Bethe-Salpeter equation. Using data from the C2DB database, simple descriptors such as layer thickness, chemical composition, and the PBE band gap were selected to capture trends in reduced screening and anisotropic dielectric response. Various ML models were trained, including Random Forest (RF), Gradient Boosting, Support Vector Machines, Neural Networks, and Kernel Ridge Regression. Among these, the RF model showed the highest accuracy, achieving $R^2 = 0.98$ for G$_0$W$_0$ band gaps and $R^2 = 0.84$ for EBEs. The study also employed Bayesian optimization to efficiently identify materials with the largest EBEs, further enhancing the discovery process \cite{javed2025machine}.

\noindent \hl{These studies show that ML can reduce data requirements through physically meaningful descriptors, transfer learning, and Bayesian optimization. However, reliable ML models for XC functional development still require datasets that include diverse bonding motifs, defects, magnetic configurations, strain, heterostructures, and different dielectric environments. The creation of high-fidelity datasets using hybrid or meta-GGA functionals could help train more generalizable models, while GW/BSE data can provide reference targets for excited-state and many-body effects} \cite{nair2025materials_database}.
\subsection{Integration with first-principles frameworks}
\noindent \hl{ML approaches are most useful when they are integrated with first-principles methods rather than used as independent black-box replacements. In this direction, ML can act as a correction to a baseline functional, a surrogate for expensive many-body calculations, or a component of self-consistent DFT workflows. Such integration is important because the model must remain compatible with the variational and self-consistent nature of Kohn-Sham DFT.}

\begin{figure}[h!]
    \centering
    \includegraphics[scale=1]{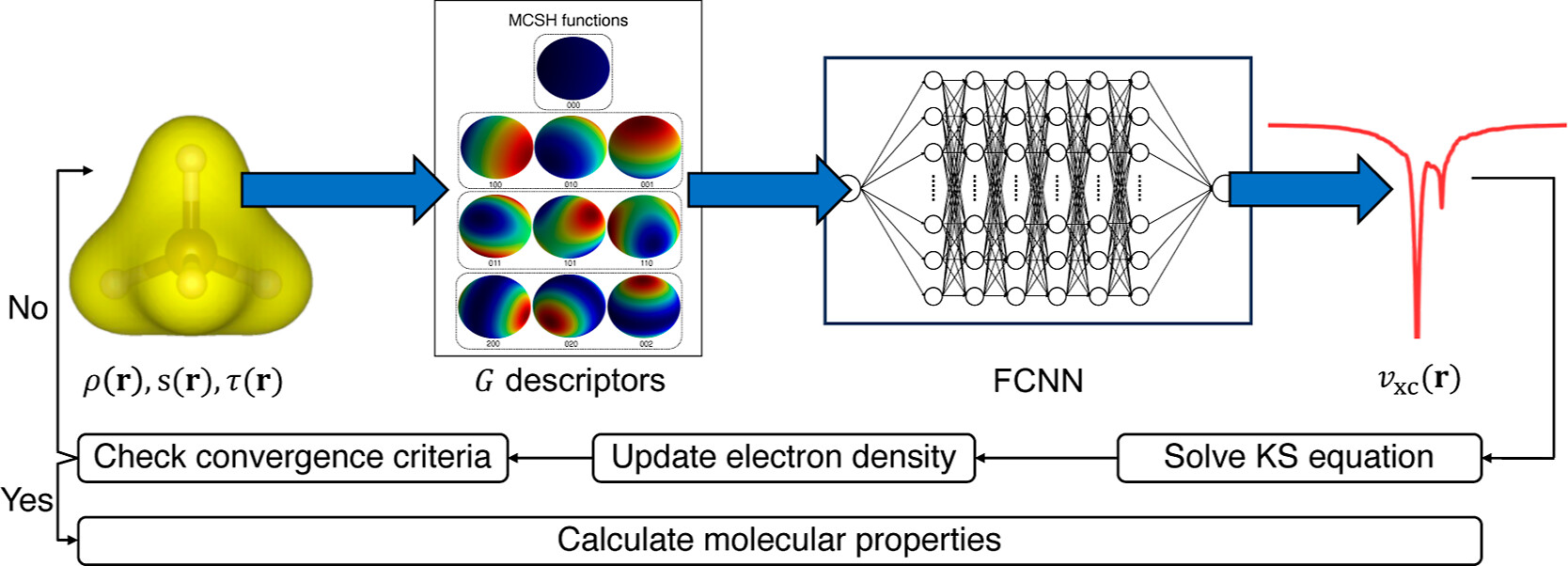}
    \caption{Schematic illustration of the self-consistent field Kohn--Sham/fully connected neural network scheme (KS-DFT/FCNN). Reprinted from \cite{zhuang2025machine}. \href{https://creativecommons.org/licenses/by-nc-nd/4.0/}{\underline{CC BY-NC-ND 4.0}}.}
    \label{fig:HRNN}
\end{figure}

\noindent Zhuang et al. (2025) proposed a ML based model to develop accurate XC potentials and reducing of delocalization errors along the way, which are common in traditional DFT functionals. Their approach utilizes a fully connected neural network (FCNN) that is trained on high-level quantum chemistry data to directly map electron density to the corresponding XC potential as shown in Fig. \ref{fig:HRNN}. By learning these complex non-local interactions, the model significantly improves the description of electron localization and binding energies. This breakthrough demonstrates the power of ML in enhancing the accuracy of XC potentials and offering a more reliable tool for DFT simulations \cite{zhuang2025machine}.

\noindent \hl{Beyond ground-state properties, ML can capture many body effects that are computationally expensive to calculate directly. For example, the EBE model discussed above uses PBE-derived and structural descriptors to approximate trends that would otherwise require GW+BSE calculations. Similarly, graph-based transfer learning has been used to climb different levels of theoretical accuracy for optoelectronic properties, where a model is first trained on lower-level data and then fine-tuned on higher-level spectra} \cite{grunert2025machine}. These approaches suggest that ML can bridge different levels of theory and reduce the number of expensive calculations required for screening 2D materials.

\noindent \hl{Machine-learning interatomic potentials (MLIPs) provide another route for integrating ML with first-principles calculations. They learn potential energy surfaces from DFT energies and forces, enabling large-scale molecular dynamics with near-ab initio accuracy and much lower computational cost} \cite{deringer2019machine}. \hl{For 2D materials, this is useful for defects, strain, and thermal transport, but the predictions still depend on the XC functional used to generate the training data.}

\subsection{Key bottlenecks and future research strategies}
\noindent \hl{Despite these advances, several bottlenecks remain. First, ML models can extrapolate poorly when applied to materials outside their training domain. This is critical for 2D systems because new materials may contain unusual bonding, strong SOC, magnetism, defects, strain, or heterostructure environments. Second, reference data for 2D materials are still limited, particularly for excitonic effects, magnetic anisotropy, phonons, and vdW heterostructures. Third, many ML models improve numerical accuracy but do not provide physically interpretable corrections to the XC functional. Finally, integration into self-consistent DFT remains challenging because the learned energy, potential, and functional derivative must be mutually consistent.}

\noindent Future research should therefore focus on physically constrained and uncertainty-aware ML models. Exact constraints, such as density scaling, spin scaling, and known bounds, should be embedded directly into ML-XC models to reduce unphysical extrapolation \cite{pokharel2022exact}. \hl{Active learning and Bayesian optimization can identify where new high-level calculations are most needed, reducing the cost of building reference datasets} \cite{javed2025machine}. \hl{Multi-fidelity and transfer-learning strategies can combine inexpensive PBE-level data with smaller sets of hybrid, GW/BSE, or experimental data} \cite{gupta2024structure, grunert2025machine}. \\ 
\noindent \hl{In the case of MLIPs, a key strategy is to train on diverse and functional-consistent datasets that include strained structures, defects, surfaces, and finite-temperature configurations. Universal graph-based potentials such as M3GNet show that broad chemical coverage can accelerate structural relaxation and molecular dynamics across materials space, but for 2D materials their reliability should still be benchmarked against high-level DFT or beyond-DFT data} \cite{chen2022universal}. \\ 
\noindent Future descriptors should explicitly capture anisotropic screening, layer thickness, dielectric environment, vdW coupling, spin-orbit effects, and defect chemistry. In this way, ML can move beyond fast property prediction and become a practical route for developing transferable, physically interpretable, and first-principles-compatible XC functionals for low-dimensional materials.

\section{Conclusion and future perspectives}
This review highlights that the performance of XC functionals in 2D materials is fundamentally property dependent. The reduced dimensionality of these systems weakens dielectric screening and enhances nonlocal exchange, electron--hole attraction, vdW interactions, and spin-orbit effects. As a result, the errors of a given functional are not universal: a functional that gives reliable lattice constants or magnetic exchange may still fail for quasiparticle gaps or excitonic spectra. Therefore, the central conclusion of this review is that \hl{no single exchange-correlation functional universally performs well for all properties of 2D materials. Optimal accuracy requires balancing physical fidelity against computational cost, guided by the property of interest and the specific material's dimensionality, bonding character, and many-body effects.}

\subsection{Functional selection framework}
\noindent \hl{Based on the reviewed literature, we propose the following practical framework for selecting the appropriate level of theory in 2D materials:
\begin{itemize}
\item \textbf{Structural and vdW-dominated properties:} SCAN/r$^2$SCAN, PBEsol, or vdW-corrected functionals are suitable starting points when lattice constants, interlayer binding, and structural relaxation are the primary targets. Their success arises from improved treatment of intermediate-range exchange and bonding, but long-range nonlocal correlations must be included for layered or heterostructured systems.
\item \textbf{Band gaps and quasiparticle energies:} HSE06 generally improves over semilocal functionals by including partial exact exchange, but its fixed screening can be insufficient for materials with strong SOC, gapless materials, or highly anisotropic dielectric response. For quantitative quasiparticle gaps, G$_0$W$_0$ or self-consistent GW remains the more physically justified approach because it explicitly includes the nonlocal, energy-dependent self-energy.
\item \textbf{Optical and excitonic properties:} GW+BSE is required when reduced screening produces strong electron-hole attraction and large EBEs. Semilocal and hybrid functionals can provide useful trends, but they do not explicitly describe bound excitons and therefore cannot reliably predict optical spectra in strongly confined 2D semiconductors.
\item \textbf{Magnetic properties:} SCAN/r$^2$SCAN are promising for magnetic exchange and transition-temperature trends because they reduce self-interaction errors and improve localization compared with GGA. However, SOC, magnetic anisotropy, and low-dimensional spin fluctuations must be treated carefully, especially for intrinsic 2D magnets.
\item \textbf{Thermal and phonon properties:} PBE and PBEsol remain efficient choices for phonons and thermal transport in many systems, but higher-order interatomic force constants and phonon lifetimes can be sensitive to the XC functional. Meta-GGA and vdW-aware functionals should therefore be benchmarked when anharmonicity, weak interlayer forces, or phase stability are important.
\item \textbf{Large-scale or high-throughput studies:} ML models and MLIPs can accelerate screening, molecular dynamics, defect modeling, and thermal simulations, but they inherit the accuracy of the reference data. Their use should therefore be tied to reliable DFT, hybrid, GW/BSE, QMC, or experimental benchmarks, depending on the target property.
\end{itemize}}

\subsection{Future perspectives}
\noindent \hl{Future progress should focus on developing XC functionals and computational workflows that are explicitly aware of low-dimensional screening. Dielectric-dependent hybrids, range-separated functionals, and nonlocal correlation corrections are promising directions because they can adapt the amount of exchange and correlation to the local environment rather than relying on fixed empirical parameters. Such approaches are especially important for heterostructures, moiré systems, defects, and interfaces, where the screening length and bonding character vary spatially.}

\noindent \hl{Another important direction is the construction of high-fidelity benchmark datasets for 2D materials. These datasets should include structural, electronic, magnetic, phononic, and excitonic properties computed using carefully converged hybrid, GW/BSE, QMC, or experimental references. They would allow systematic assessment of functional transferability and provide reliable targets for ML-XC models and MLIPs. In parallel, uncertainty-aware ML, active learning, and multi-fidelity training can reduce the cost of generating reference data while identifying where more accurate calculations are most needed.}

\noindent \hl{In summary, the future of predictive modeling for 2D materials lies not in a single universal functional, but in a hierarchy of methods chosen according to the relevant length scale, screening regime, and target property. A physically guided combination of semilocal/meta-GGA functionals, hybrid and many-body methods, pseudopotential consistency, and ML acceleration offers the most practical route toward accurate and computationally efficient design of 2D materials.}


\section*{Acknowledgement} This work is supported by the National Natural Science Foundation of China (Grant No. 12204037). We also acknowledge
the support from the Beijing Institute of Technology Research Fund Program for Young Scholars.


\data{The data that support the findings of this study are available from the corresponding author upon reasonable request.}

\section*{Conflict of Interest}
The authors declare no competing financial interest.
\bibliographystyle{iopart-num}
\bibliography{reference}

\end{document}